\documentclass[12pt]{article}

\usepackage{amsmath, amssymb}
\usepackage[]{graphicx}
\usepackage{subeqnarray}
\newcommand{\<}{\langle}
\newcommand{\FT}{\Phi^\times}
\newcommand{\F}{{\Phi}}

\newcommand{\G}{\Gamma}
\newcommand{\HH}{{\cal H}}

\newcommand{\rhsp}{\F_+\subset\HH\subset\F^\times_+}
\newcommand{\rhsm}{\F_-\subset\HH\subset\F^\times_-}
\newcommand{\Om}{\Omega}
\newcommand{\om}{\omega}
\newcommand{\R}{{\rm {I\!R}}}

\newcommand{\C}{{\rm {l\!\!\! {C}}}}

\newcommand{\CM}{{\cal M}}

\newcommand{\CS}{{\cal S}}
\newcommand{\CN}{{\cal N}}
\newcommand{\CH}{{\cal H}}

\newcommand{\bee}{\begin{enumerate}}
\newcommand{\ene}{\end{enumerate}}
\newcommand{\een}{\end{enumerate}}
\newcommand{\bea}{\protect{\begin{align}}}
\newcommand{\eea}{\protect{\end{align}}}
\newcommand{\beq}{\protect{\begin{equation}}}
\newcommand{\eeq}{\protect{\end{equation}}}

\newcommand{\klam}{{\!\text{{\Large $\prec$}}\!\,}}
\newcommand{\mer}{{\!\,\text{{\Large $\succ$}}\!}}
	
\begin{document}

\def\theequation{\thesection.\arabic{equation}}
\setcounter{equation}{0}

\title{ \vspace{-.5in}{\bf Gamow-Jordan Vectors and Non-Reducible
Density Operators from Higher Order S-Matrix Poles}} \vspace{4pt}
\author{
	{\bf A. Bohm}\thanks{Center for Particle Physics}, 
 	{\bf M. Loewe}\thanks{Microelectronics Research Center},
			\vspace{8pt}\\
	{\bf S. Maxson}\thanks{current address: Department of Physics,
		University of Colorado at Denver, Denver, Colorado 80217-3364},
	{\bf P. Patuleanu}$^*$,
	{\bf C. P{\"u}ntmann}$^*$  \vspace{4pt}\\ 
		\footnotesize{The University of Texas at Austin}\\
		\footnotesize{Austin, Texas~78712}
			\vspace{4pt}\\
			    and
			\vspace{4pt}\\
	{\bf M. Gadella} \vspace{4pt}\\
		\footnotesize{Faculdad de Ciencias, 
			      Universidad de Valladolid}\\
		\footnotesize{E-47011 Valladolid, Spain}
}
\maketitle

\begin{abstract}
In analogy to Gamow vectors that are obtained from first order
resonance poles of the S-matrix, one can also define higher order Gamow
vectors which are derived from higher order poles of the S-matrix. An
S-matrix pole of $r$-th order at $z_R=E_R-i\Gamma/2$ 
leads to $r$ generalized eigenvectors of order $k=0,~1, \ldots,~r-1$,
which are also Jordan vectors of degree $(k+1)$ with generalized eigenvalue
$(E_R-i\Gamma/2)$. The Gamow-Jordan vectors are elements of a
generalized complex eigenvector expansion, whose form suggests the
definition of a state operator (density matrix) for the microphysical
decaying state of this higher order pole. 
This microphysical state is a mixture of
non-reducible components. In spite of the fact that the $k$-th order
Gamow-Jordan vectors has the polynomial time-dependence which one
always associates with higher order poles, 
the microphysical state obeys a purely exponential decay law.
\end{abstract}

\newpage
\section{Introduction}\label{sec:introduction}
\setcounter{equation}{0}

The singularities of the analytically continued S-matrix that have
attracted most of the attention in the past are the first order poles
in the second sheet. They were associated with resonances that decay
exponentially in time.~\cite{doublepoles} In conventional Hilbert space
quantum theory it was not clear what those resonance ``states'' 
were, since a vector description of a resonance state was not possible
within the framework of the Hilbert space.\cite{fonda-ghirardi-rimini} 
Higher order poles, in particular double poles have also been 
mentioned, but it has long been believed that they
somehow lead to an additional polynomial time dependence of the decay
law~\cite{polynomial}. However, precise derivations were not possible
due to the lack of a vector space description.

This changed when the first order poles were associated with vectors 
{${\psi^G =\sqrt{2\pi\G}|E_R-i\Gamma/2\rangle}$} in a
rigged Hilbert space (RHS)~\cite{bohm, bohm-group, gadella1983, 
bohm-gadella}, called Gamow vectors. They possess
all the properties that one needs to describe decaying states or 
resonances: These Gamow vectors $\psi^G$ are eigenvectors of a 
self-adjoint Hamiltonian~\cite{22a} with complex eigenvalues
$z_R=E_R-i\Gamma/2$ (energy and width). They evolve
exponentially in time, and they have a Breit-Wigner energy
distribution. They also obey an exact Golden Rule, which becomes the
standard Golden Rule if one replaces $\psi^G$ with its Born
approximation. The existence of these vectors allows us to 
interpret resonances as autonomous
physical systems (which one cannot do in standard quantum
mechanics). It also puts quasibound states (i.e. resonances) and
anti-bound (or virtual) states~\cite{gadella1983} on the same 
footing with the bound states (eigenvectors with real energy), 
which have both a vector description and 
an S-matrix description. Mathematically, 
Gamow vectors are a generalization of Dirac kets (describing 
scattering states), i.e. they are also eigenkets. But whereas 
Dirac kets are associated with a value of the continuous Hilbert space
spectrum of the self-adjoint Hamiltonian $H$, the Gamow kets are not,
but have complex eigenvalues.

Using the entirely different theory of finite dimensional complex
matrices, decaying states (like the $K^0-\bar{K}^0$ system) have been
phenomenologically described as eigenvectors of an effective
Hamiltonian matrix with complex eigenvalues.  One usually assumes that
these complex Hamiltonians are diagonalizable~\cite{lee}. However, unlike
hermitean matrices which have real eigenvalues, non-hermitean 
finite dimensional matrices cannot always be diagonalized, but
can only be brought into a Jordan canonical
form~\cite{baumgartel-gantmacher-lancaster}. Finite dimensional
matrices consisting of non-diagonalizable Jordan blocks
have been mentioned in connection with resonances numerous times in
the past~\cite{lukierski,brandas,antoniou-tasaki,mondragon,stodolsky},
and they have been used for discussions of problems in
nuclear~\cite{mondragon} and in hadron~\cite{stodolsky} physics. 
Jordan blocks have also been obtained in
prototypes of mixing systems~\cite{antoniou-tasaki}, 
and the appearance of so-called
``irreducible'' non-diagonalizable blocks in the density matrix has
been sought after for some time in connection with irreversible
thermodynamics and the approach to equilibrium~\cite{prigogine}. 
That irreducible non-diagonalizable Jordan blocks may shed
light on the idea of quantum chaos has been mentioned by Br{\"a}ndas and
Dreismann~\cite{brandas}. Also important for the understanding of
quantum chaos and of the statistical properties of nuclear spectra are
accidental degeneracies and level crossing, which in the past had been
almost exclusively restricted to stable systems driven by hermitean
Hamiltonians~\cite{wilczek}. Based on a finite dimensional
phenomenological expression for the S-matrix~\cite{verbaarschot},
Mondrag\'{o}n {\em et al.}~\cite{mondragon} extended these discussions
to resonance states described by a Jordan block of rank $2$. 

In the present paper we shall show that the Jordan blocks emerge
naturally for the matrix elements
$\,^{(k)}\langle^-z_R|H|\psi^-\rangle$ of a self-adjoint~\cite{22a}
Hamiltonian $H$ between Gamow vectors $|z^-_R\rangle^{(k)}$ of order
$k=1,~2,\dots,r-1$. From the generalized basis vector expansion
derived here it follows that these $r$-dimensional blocks are a
truncation of the infinite dimensional exact theory in the RHS. 

The higher order Gamow vectors $|z^-_R\rangle^{(k)}$ have been derived
in a recent unpublished preprint by Antoniou and
Gadella~\cite{antoniou-gadella}. The derivation is a 
generalization of the method by which the Gamow vector 
(of order $k=0$) was derived from the first 
order poles of the S-matrix~\cite{bohm-group}.  
Starting from an $r$-th order pole of the S-matrix 
element at complex energy $z=z_R$, they derived $r$ 
Gamow vectors of higher order, $|E_R-i\Gamma/2\,^-\rangle^{(k)}$, 
$k=0,\, 1,\cdots r-1$, as functionals in
a rigged Hilbert space. These higher order Gamow kets
are also Jordan vectors belonging to the eigenvalue $z_R$.  

In the present paper we generalize the RHS theory of the Gamow vectors
associated with first order S-matrix poles, which we call Gamow vectors
of order zero, to poles of order $r$. Quasistationary states in
scattering experiments (i.e. states formed if the projectile is
temporarily captured by the target) can be shown to appear not only as
first order poles, but as poles of any order $r=1,2,\dots$
(\cite{polynomial}). In section~\ref{sec:poles}, 
we will start from the expression for the unitary S-matrix of a
quasistationary state of finite order $r$ and energy $E_R$, given in
reference~\cite{bohm} sect.~XVIII.6, and obtain from it $r$ Gamow
vectors of order $k=0,1,\dots,r-1$ which are also Jordan vectors of
degree $k+1$. After a review of the case $r=1$ in
section~\ref{sec:summary}, we derive in section~\ref{sec:higher} 
the generalized eigenvector expansion, which contains the Gamow-Jordan
vectors as basis vectors. With these basis vectors we can give a
matrix representation of $H$ and of $e^{-iHt}$ which contains the
$r$-dimensional Jordan blocks. In section~\ref{sec:possible}, we start
from the pole term of the $r$-th order S-matrix pole and conjecture the state
operator for the hypothetical microphysical system associated with
this pole. This $r$-th order Gamow state operator consists of
non-diagonalizable blocks which obey a purely exponential decay law.
This unexpected result is in contrast to the belief~\cite{polynomial}
that higher order poles must lead to an additional polynomial time dependence.

At the present time there is little empirical evidence for the existence
of these higher order pole ``states'' in nature. This is in marked
contrast to the fact that first order pole states described by
ordinary Gamow vectors have been identified in abundance, e.g. through
their Breit-Wigner profile in scattering experiments and through their
exponential decay law. 

Now that our results have obliterated the prime empirical objection of
non-exponentiality against the existence of higher order pole states,
one can continue to look for them. The first step in this direction is
to use these higher order state operators in the exact Golden
Rule~\cite{bohm} and obtain the decay probability and the decay rate,
including the line widths. We plan to do this in a forthcoming paper.

\section{Poles of the S-matrix and Gamow-Jordan Vectors}\label{sec:poles} 
\setcounter{equation}{0}

Since the new (hypothetical) states are to be defined by the $r$-th
order pole of the S-matrix, we consider a scattering system. The
S-matrix consists of the matrix elements~\cite{s-matrix}
\begin{align}
	\hspace{-.1in}
	\left(\psi^{\text{out}},\phi^{\text{out}}\right)&=
      	\left(\psi^{\text{out}}(t),
        \phi^{\text{out}}(t)\right)
        =\left( \psi^{\text{out}},S \phi^{\text{in}}\right) 
\nonumber  \\ \vspace{6pt}
  	& =  \left( \Om^- \psi^{\text{out}}(t), \Om^+ \phi^{\text{in}}(t)
        \right)    
    	= \left( \psi^-(t),\phi^+(t)\right) = \left( \psi^-,\phi^+\right) 
\label{eq:outmatrix} \\
  	& =\int_{\text{spectrum }H}dE\;\langle\psi^-|E^-\rangle 
        S(E+i0)\langle^+  E|\phi^+\rangle\;.\nonumber
\end{align}
Since we are interested only in the principles here, in equation
(\ref{eq:outmatrix}) (and in subsequent equations) we choose to
ignore all other labels of the basis vectors $|E^\pm \rangle$ and
$|E\rangle$ except the energy label $E$, which can take values on a
two-sheeted Riemann surface.  Nothing principally new will be gained
if we retain the additional quantum numbers $b= b_2,\, b_3, \ldots
b_\CN$ in the basis system $|E^\pm \rangle \Longrightarrow |E,
b^\pm\rangle = |E,\, b_2,\, b_3,\,\ldots b^\pm_\CN\rangle$, and in place
of the integral over the energy we would just have some additional
sums (or integrals in the case that some of the $b$'s are continuous)
over the quantum numbers $b$.  For instance, if one chooses the angular
momentum basis $|E^\pm\rangle \Longrightarrow |E,l,\, l_3,\, \eta^\pm
\rangle$, where $\eta$ are some additional (polarization or) channel
quantum numbers (cf.~\cite{bohm}, sect.~ XX.2, XXI.4), 
then~(\ref{eq:outmatrix}) would read in detail
\begin{align}
	\left( \psi^- ,\phi^+\right) =& \sum_{l,\, l_3,\, \eta} \sum_{l^\prime
	,\, l^\prime_3 ,\, \eta^\prime} \iint\, dE\, dE^\prime \, \langle
	\psi^-\, |\, E^\prime ,\, l^\prime ,\, l^\prime_3 ,\, \eta^{\prime -}
	\rangle  \times 
\label{eq:matrix2}\\
	&\quad \times\langle^- E^\prime ,\, l^\prime ,\, l^\prime_3 ,\,
	\eta^\prime \, |E,\, l,\, l_3 ,\, \eta^+\rangle\langle^+ E,\,
	l,\, l_3 ,\, \eta\, |\, \phi^+\rangle \; . \nonumber
\end{align}
Restricting ourselves to one initial $\eta =\eta_A$ and one final
$\eta'=\eta_B$ channel (e.g., $\eta_B=\eta_A$ for elastic
scattering) we obtain
\begin{eqnarray}\label{eq:matrix2a}
	\langle^-E',l',l'_3 ,\eta_B|E,l,l_3,\eta_A^+\rangle &=&
	\langle E',l',l'_3,\eta_B|\,S\,|E,l,l_3,\eta_A\rangle\\ 
	&=&\delta(E'-E)\delta_{l^\prime_3l_3}\delta_{l^\prime l}
	\langle\eta_B|\!|S|\!|\eta_A\rangle\nonumber
\end{eqnarray}
where
\begin{equation}
	\langle \eta_B \, |\!| \, S\, |\!| \, \eta_A\rangle =
S^{\eta_B}_l (E)
\label{eq:matrix2b}
\end{equation}
is the $l$-th partial S-matrix element for scattering from the channel
$\eta_A$ into one particular channel $\eta_B$ (e.g., the elastic channel,
$\eta_B=\eta_A$).  If we consider the $l$-th partial wave of the
$\eta_B$-th channel, then the $S(E)$ in (\ref{eq:outmatrix}) is given
by this matrix element $S(E)=S^{\eta_B}_l (E)$.  E.g., if we consider
a mass point in a potential barrier, then $|E^\pm\rangle = |E,\, l,\,
l^\pm_3 \rangle$ is the angular momentum basis of the mass point and,
depending on the shape and height of the barrier, one or several
resonances can exist. Many concrete examples have been studied where
one can see how first order resonance poles $z_{R_i}=E_{R_i} -i\, \G_i
/2$ move as a function of the potential parameters~\cite{hogreve}.  
We want to consider just one pole, and in the
present paper we are mainly interested in a higher order pole at
$z_R$.  Whether physical systems exist that are described by higher
order poles is not clear, but a few examples of second order poles
have been discussed in the past~\cite{polynomial}~\cite{mondragon}.

With the above simplifications to one channel $\eta_B$ and one partial
wave $l$, the notation in (\ref{eq:outmatrix}) is standard in
scattering theory. 
\begin{figure}[t!]
\includegraphics[]{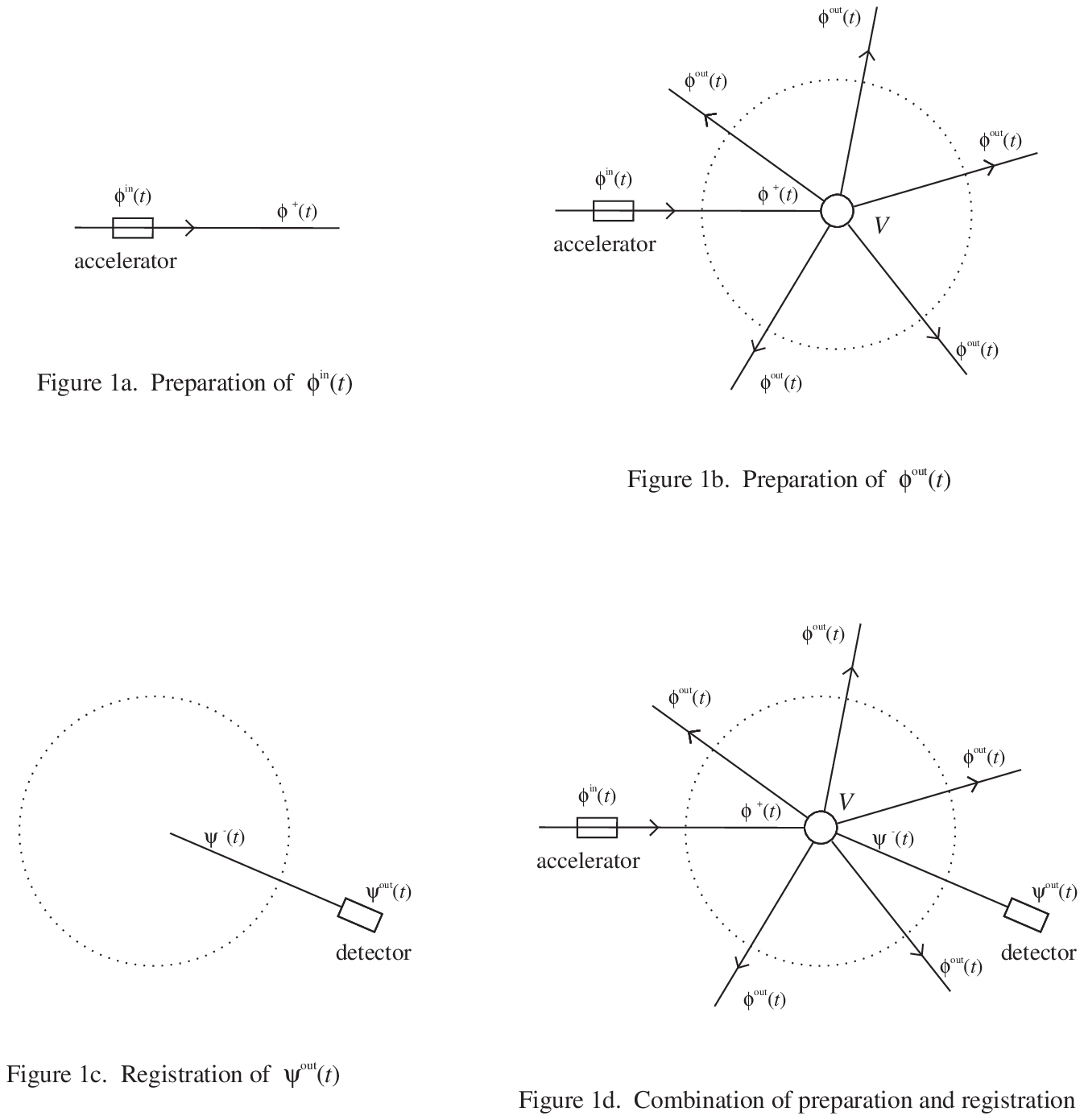}
\caption{The preparation-registration procedure in a scattering experiment}
\end{figure}
The standard scattering theory uses the same Hilbert space ${\cal H}$
for both the set of in-states $\phi^+$ and the set of out-``states''
$\psi^-$. The RHS formulation allows us to use two RHS's for the set
$\{\phi^+\}$ defined by the initial conditions and the set $\{\psi^-\}$
defined by the final conditions. To explain this we subdivide the
scattering experiment into a preparation stage and a registration
stage, as explained in detail in reference~\cite{cologne}.
Fig.~1 depicts these different stages illustrating the idealized process:  
The in-state $\phi^+$ (precisely the state which evolves from the
prepared in-state $\phi^{\text{in}}$ outside the interaction region 
where $V=H-H_0$ is zero) is determined by the accelerator. The so-called
out-state $\psi^-$ (or $\psi^{\text{out}}$) is determined by the 
detector; $|\psi^{\text{out}} \rangle \langle\psi^{\text{out}} |$ 
is therefore the observable which the detector registers and not 
a state. In the conventional formulation one describes both 
the $\phi^{\text{in}}$ and the $\psi^{\text{out}}$ by
any vectors of the Hilbert space. In reality the $\phi^{\text{in}}$ 
(or $\phi^+$) and $\psi^{\text{out}}$ (or $\psi^-$) are subject 
to different initial and boundary conditions and are therefore 
described by different sets of vectors belonging to different 
rigged Hilbert spaces. The RHS for the Dirac kets is denoted by
\begin{equation}\label{eq:rhs}
	\Phi \subset \CH \subset \Phi^\times
\end{equation}
where $\Phi$ is the space of the ``well-behaved'' vectors (Schwartz
space), and the Dirac kets (scattering states) $|E^\pm\rangle$ and  
$|E\rangle$ are elements of $\Phi^\times$.
The in-state vectors $\phi^+(t)=e^{iHt/\hbar}\phi^+$ evolve from 
the prepared in-state $\phi^{\text{in}}(t)=(\Om^+)^{-1}\,\phi^+(t),\,
t<0,$ and the out-observable vectors $\psi^-(t)=e^{iHt/\hbar}\psi^-$ 
evolve into the measured out-state $\psi^{\text{out}}(t)=(\Om^-)^{-1}\,
\psi^-(t),\, t>0$.\cite{fussnote}

We denote the space of $\{\phi^+\}$ by $\F_-$ and the space of $\{\psi^-\}$ by
$\F_+$.  Then, $\F = \F_- +\F_+$, where $\F_-\cap\F_+ \ne \emptyset$. In
place of the single rigged Hilbert space (\ref{eq:rhs}), one therefore has a
pair of rigged Hilbert spaces:
\begin{subeqnarray}\label{eq:rhsm}\label{bohm6}
	\phi^+\in\; \rhsm&&
	\text{for in-states of a scattering}\\\nonumber
	&&\text{experiment which are prepared}	\\\nonumber
	&&\text{by a preparation apparatus,}\\
	\psi^-\in\; \rhsp&&			\label{bohm7}
	\text{for observables or out-``states''}\\\nonumber
	&&\text{which are registered by a detector.}
\end{subeqnarray}
The Hilbert space ${\cal H}$ in (\ref{eq:rhs}), (\ref{bohm6}a), and 
(\ref{bohm7}b) is the same, but $\F_+$ and $\F_-$ are two distinct
spaces of ``very well-behaved'' vectors. The spaces $\F_+$ and $\F_-$
can be defined mathematically in terms of the spaces of their wave
functions $\langle^+ E|\phi^+\rangle$ and $\langle^- E|\psi^-\rangle$,
respectively.  This is the realization of these abstract spaces by
spaces of functions, in very much the same way as the Hilbert space
${\cal H}$ is realized by the space of 
{L}{e}{b}{e}{s}gue square-integrable functions
{$L^2 [0,\infty )$}. The space $\F_-$ is realized by the space of 
well-behaved Hardy class functions in the lower half-plane of the
second energy sheet of the S-matrix $S(E)$, and the space $\F_+$ is
realized by the space of well-behaved Hardy class functions in the upper
half-plane. Thus, the mathematical definition of the spaces $\Phi_+$
and $\Phi_-$ is:  
\begin{equation}
\psi^-\in\F_+\quad\text{iff}\quad
\langle\,E\,|\,\psi^{\text{out}}\rangle =\langle^- E\, 
  |\psi^-\rangle \in \CS\cap\HH^2_+ \, 
     \Bigr|_{\R^+} 
\label{eq:2.11a}
\end{equation}
and
\begin{equation}
\phi^+\in\Phi_- \quad \text{iff}\quad
\langle\,E\,|\,\phi^{\text{in}}\rangle =\langle^+ E \,
   |\phi^+\rangle \in \CS\cap\HH^2_-\, 
     \Bigr|_{\R^+} \;.
\label{eq:2.11b}
\end{equation}
Here ${\cal S}$ denotes the Schwartz space and $\CS\cap\HH^2_\pm$ is
the space of Hardy class functions from above/below. This mathematical
property of the spaces $\F_+$ and $\F_-$ can be shown to be a
consequence of the arrow of time inherent in every scattering
experiment~\cite{cologne}. 

Being Hardy class from below means that the analytic continuation
$\langle\psi^-|z^-\rangle$ of $\langle \psi^-|E^-\rangle =
\overline{\langle^- E|\psi^-\rangle}$, and the analytic continuation
$\langle^+z|\phi^+\rangle$ of $\langle^+ E|\phi^+\rangle$, and therewith also
$\langle\psi^-|z^-\rangle\langle^+z|\phi^+\rangle$, are analytic
functions in the lower half-plane which vanish fast enough on the
lower infinite semicircle.  (For the precise definition,
see~\cite{bohm-gadella, duren-hoffman}). The values of a Hardy class
function in the lower half-plane are already determined by its values on the positive real
axis~\cite{vanwinter}. From (\ref{eq:2.11a}) and (\ref{eq:2.11b}) follows that
\begin{equation}\label{eq:2.12}
	\langle\psi^-|E^-\rangle\langle^+ E|\phi^+ \rangle
	\in\CS\cap\HH^p_-\;, \; p=1
\end{equation}
and so are all its derivatives 
\begin{equation}\label{eq:2.13}
	\hspace{-.4in}\bigl( \langle\psi^-|E^-\rangle\langle^+ E|
	\phi^+ \rangle \bigr)^{(n)}\in\CS\cap\HH^p_-\;;\hspace{.2in}
	p=1,~n=0,\, 1,\, 2,\, \ldots
\end{equation}
because the derivation is continuous in $\CS$. 

With the above preparations one can derive the vectors that are
associated with the $r$-th order pole of the S-matrix for any value
of $r$, in complete
analogy to the derivation of the vectors associated with the first
order poles, $r=1$.\cite{bohm-group}~ ~ ~We shall see that there are $r$
generalized vectors of order $k=0,\, 1, \ldots ,\, r-1$ associated 
with an $r$-th order pole.  We call these vectors the higher order Gamow
vectors, or Gamow-Jordan vectors (since they also have the properties
of Jordan vectors~\cite{baumgartel-gantmacher-lancaster}). 
Their first derivation from the $r$-th order pole was given 
in~\cite{antoniou-gadella}.  Here we give an alternative
derivation and discuss their properties and applications in the
generalized basis vector expansion.  

We consider the model in which the analytically continued S-matrix $S(\omega)$
has one $r$-th order pole at the position $\omega=z_R$ ($z_R =E_R-i\,
\G /2$) in the lower half-plane of the second sheet, (and consequently
there is also one $r$-th order pole in the upper half-plane of the
second sheet at $\omega=z^\ast_R$). In this paper we will not discuss
the pole at $z^\ast_R$. It leads to $r$ growing higher order Gamow vectors 
and the correspondence between the growing and decaying vectors is
just the same as for the case $r=1$. The model that we discuss here 
can easily be extended to any finite number of finite order poles in
the second sheet below the positive real axis.
\begin{figure}[t!]
\includegraphics*[scale=0.6,bb= 86 528 372 722, clip]{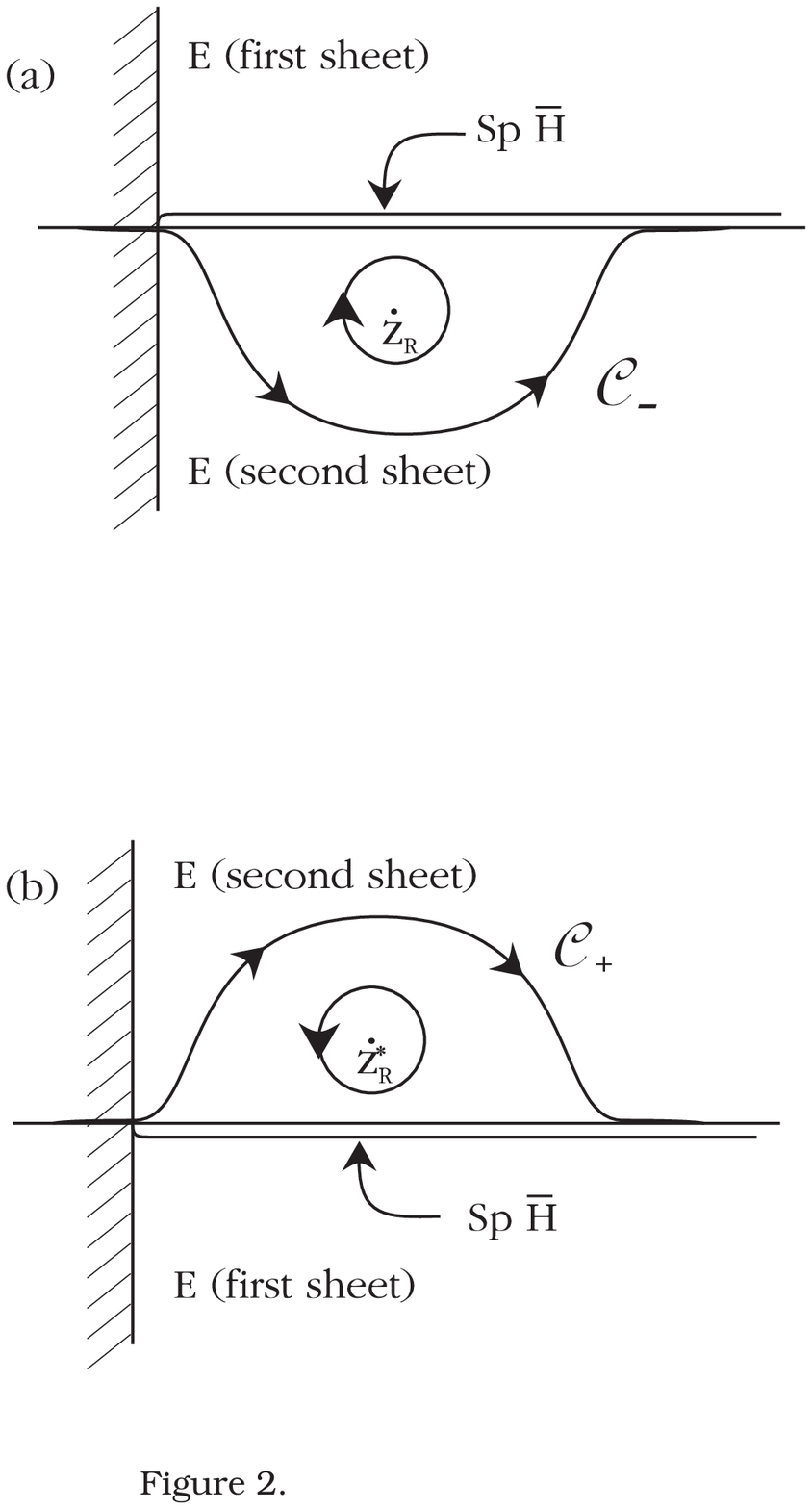}\hspace{35pt}
\includegraphics*[scale=0.6,bb= 86 210 372 422, clip]{sub2.eps}
	\caption{The contours in the two sheeted Riemann surface. 
	(a) displays the contour ${\cal C}^-$ that results from 
	extending the spectrum of the Hamiltonian 
	${\rm Sp}(\bar{H}) = {\rm{I\!R}}^+$ into the lower half-plane 
	of the second Riemann sheet and that yields the 
	pole term in eq.~(\ref{paul17}) at the position 
	$z_R=(E_R-i\Gamma/2)$. (b) displays the extension of the
	contour into the upper half-plane of the second sheet with
	pole at $z^*_R$, which we shall not discuss here any further; it leads
	to the growing higher order Gamow vectors.} 
\end{figure}
		%

The unitary S-matrix of a quasistationary state associated with an
$r$-th order pole at $z_R=E_R-i\Gamma/2$ is represented in the
lower half-plane of the second sheet by (\cite{bohm} sect.~XVIII.6)
\begin{eqnarray}\nonumber
	S_{\rm II}(\omega)&=&e^{2ir\;{\rm arctan}(\frac{\Gamma}
	{2(E_R-\omega)})}e^{2i\gamma(\omega)}=\left(\frac{\omega-E_R-i\Gamma/2}
	{\omega-(E_R-i\Gamma/2)}\right)^re^{2i\gamma(\omega)}\\
	&=&\left(1+\frac{-i\Gamma}{\omega-(E_R-i\Gamma/2)}\right)^r
	e^{2i\gamma(\omega)}\;.					\label{b16a}
\end{eqnarray}
Here, $\delta_R(\omega)=2ir\,{\rm arctan}(\frac{\Gamma}{2(E_R-\omega)})$
is the rapidly varying resonant part of the phase shift, and
$\gamma(\omega)$ is the background phase 
shift, which is a slowly varying function of the complex energy
$\omega$. We have restricted ourselves to the case that the $r$-th
order pole is the only singularity of the S-matrix. Below we will
mention how this can be generalized to the case of a finite number of
finite order poles. (Note that phenomenologically only a finite number
of first order poles have been established, but there is no
theoretical reason that would prevent other isolated singularities on
the second sheet below the real axis.)

For our calculations we have to write~(\ref{b16a}) in the form of
a Laurent series:
\begin{eqnarray}\label{b16b}
	S_{\rm II}(\omega)&=&\sum_{l=-r}^{+\infty} C_l(\omega-z_R)^l\\
	&=&\frac{C_{-r}}{(\omega-z_R)^r}+\frac{C_{-r+1}}{(\omega-z_R)^{r-1}}
	+\dots+C_0+C_1(\omega-z_R)+\dots\nonumber
\end{eqnarray}
Therefore we expand the bracket in~(\ref{b16a}):
\begin{eqnarray}\nonumber
	S_{\rm II}(\omega)&=&\left(\sum_{l=0}^r\begin{pmatrix}r\\l\end{pmatrix}
	\frac{(-i\Gamma)^l}{(\omega-z_R)^l}\right)e^{2i\gamma(\omega)}\\
	&=&e^{2i\gamma(\omega)}+\sum_{l=1}^{r}\begin{pmatrix}r\\l\end{pmatrix}
	\frac{(-i\Gamma)^l}{(\omega-z_R)^l}e^{2i\gamma(\omega)}
	\label{b16c}\\
	&=&e^{2i\delta_R(\omega)}e^{2i\gamma(\omega)}\nonumber	
\end{eqnarray}
We insert this into~(\ref{eq:outmatrix}) and deform the contour of
integration through the cut along the spectrum of $H$ into the second sheet,
as shown in fig.~2a. Then one obtains
\begin{subeqnarray}\label{paul17}\label{bohm17}
	(\psi^-,\phi^+)&=&\int_{{\cal C}_-}\,d\om \;\langle\psi^-|\om^-
	\rangle\,S_{\rm II}(\omega)\,\langle^+\om |\phi^+\rangle+\\ 
	&&+\sum_{n=0}^{r-1}\oint_{\hookleftarrow} d\om\,
        \langle\psi^-|\om^-\rangle\,\frac{e^{2i\gamma(\omega)}a_{-n-1}}
	{(\om-z_R)^{n+1}}\,\langle^+\om|\phi^+\rangle\nonumber\\
	&=&\int_0^{-\infty_{\rm II}}dE\langle\psi^-|E^-\rangle\,S_{\rm
	II}(E)\,\langle^+E|\phi^+\rangle\,+\,(\psi^-,\phi^+)_{\rm P.T.}
\end{subeqnarray}
In here, $\text{Im} \,\om <0$ on the second sheet, and 
\begin{equation}\label{b17b}
	a_{-n-1}\equiv\begin{pmatrix}r\\n+1\end{pmatrix}(-i\Gamma)^{n+1}\;.
\end{equation}
The first integral does not depend on the pole and may be called a
``background term''. The contour ${\cal C}_-$ can be deformed into the
negative axis of the second sheet from 0 to $-\infty_{\rm II}$. We
shall set this background integral aside for the moment. For the
second term on the right-hand side of (\ref{bohm17}), the higher
order pole term $(\psi^-,\phi^+)_{\rm P.T.}$, we obtain using the
Cauchy integral formulas $\oint_{\hookrightarrow}\frac{f(\om )}{(\om
-z_R)^{n+1}}\;d\om=\frac{2\pi i}{n!}\left. f^{(n)}(z)\right|_{z=z_R}$ 
where $f^{(n)}(z)\equiv\frac{d^nf(z)}{dz^n}$:
\begin{eqnarray}\label{bohm18}
	(\psi^-,\phi^+)_{\rm P.T.}&\equiv&
	\sum_{n=0}^{r-1}\oint_{\hookleftarrow}
 	d\om\langle\psi^-|\om^-\rangle\,\frac{e^{2i\gamma(\omega)}a_{-n-1}}
	{(\om -z_R )^{n+1}}\,\langle^+\om |\phi^+ \rangle\\
	&=& \sum_{n=0}^{r-1}\left(-\frac{2\pi i}{n!}\right)\,a_{-n-1}
   	\Bigl(\langle \psi^-|\om^-\rangle\; e^{2i\gamma(\omega)}
	\;\langle^+ \om |\phi^+\rangle\Bigr)^{(n)}_{\om =z_R}\nonumber
\end{eqnarray}
In here, $\left(\dots\right)^{(n)}_{\omega=z_R}$ means the $n$-th
derivative with respect to $\omega$ taken at the value $\omega=z_R$.
Since the kets $|\omega^-\rangle$ are (like the Dirac kets
$|E^-\rangle$) only defined up to an arbitrary factor or, if their
``normalization'' is already fixed, up to a phase factor we absorb the
background phase $e^{2i\gamma(\omega)}$ into the kets
$|\omega^-\rangle$ and define new vectors
\begin{equation}\label{b18a}
	|\omega^\gamma\rangle\equiv|\omega^-\rangle 
	e^{2i\gamma(\omega)}\;.
\end{equation}
(Note that the phase is not trivial since e.g. $|E^-\rangle
e^{2i\delta(E)}=|E^+\rangle$.) For the case that the slowly varying
background phase $\gamma(\omega)$ is constant, the
$|\omega^\gamma\rangle$ are up to a totally trivial constant phase
factor identical with $|\omega^-\rangle$; but in
general~(\ref{b18a}) is a non-trivial gauge transformation. For the
case of a first order resonance pole, $n=r-1=0$ in~(\ref{bohm18}), the
phase transformation~(\ref{b18a}) is also irrelevant, because for 
$n=0$ no derivatives are involved in~(\ref{bohm18}). Using the phase
transformed vectors~(\ref{b18a}) we can proceed in the same way as
if we were using the $|\omega^-\rangle$ with $\gamma(\omega)={\rm constant}$.

Taking the derivatives, we rewrite~(\ref{bohm18}) as:
\begin{equation}\label{bohmprime}
	\hspace{-.3in}(\psi^-,\phi^+)_{\rm P.T.}=\sum_{n=0}^{r-1}\left(
	-\frac{2\pi i}{n!}a_{-n-1}\right)\sum_{k=0}^n
	\begin{pmatrix}n\\k\end{pmatrix}\langle \psi^-|
	z_R^\gamma\rangle^{(k)}\,^{(n-k)}\langle^+z_R|\phi^+\rangle 
\end{equation}
In here, we denote by $\langle\psi^-|z^\gamma\rangle^{(n)}$ the 
$n$-th derivative of the analytic function
$\langle\psi^-|z^\gamma\rangle$, and with
$\langle\psi^-|z^\gamma_R\rangle^{(n)}$ its value at $z=z_R$.
Since $\langle\psi^- |E^-\rangle \in \CS\cap\HH^2_-$, it follows that
$\langle\psi^- |z^-\rangle^{(n)}$ and $\langle\psi^-
|z^\gamma\rangle^{(n)}$ are also analytic functions in the
lower half-plane of the second sheet, whose boundary values on the positive
real axis have the property $\langle\psi^-
|E^\gamma\rangle^{(n)}\in\CS\cap\HH^2_-$.  

Analogously,  we denote by $\,^{(n)}\langle^+ z|\phi^+\rangle$ the 
$n$-th derivative of the analytic function $\langle^+ z|\phi^+\rangle$.
Again, $\,^{(n)}\langle^+ z|\phi^+\rangle$ is analytic in the lower
half-plane with its boundary value on the real axis being 
$\,^{(n)}\langle^+E|\phi^+\rangle\in\CS \cap\HH^2_-$.

The case $r=1$ (and therefore $n=0$ and $k=0$ in~(\ref{bohmprime})) is
the well-known case of the first order pole term, which led to the
definition of the ordinary Gamow vectors for Breit-Wigner resonance 
states~\cite{bohm,bohm-group,gadella1983,bohm-gadella}. We shall
review its properties in section~\ref{sec:summary}. In
section~\ref{sec:higher} we shall then discuss the general $r$-th order
pole term and the generalized vectors $|z^\gamma_R\rangle^{(k)}$,
{$k=1,~2,\dots,r-1$}. These vectors we call Gamow vectors of order $k$
or Gamow-Jordan vectors of degree $k+1$, for reasons that will become
clear in section~\ref{sec:higher}.

\section{Summary of the Case $\mathbf{r=1}$}\label{sec:summary}
\setcounter{equation}{0}

For the case $r=1$ we obtain from~(\ref{bohm18}) and~(\ref{bohmprime}): 
\begin{eqnarray}\nonumber
	(\psi^-,\phi^+)_{\rm P.T.} 
	&=&\int^{+\infty}_{-\infty_{\rm II}}dE\;\langle\psi^-|E^-\rangle
	\langle^+E|\phi^+\rangle\frac{e^{2i\gamma(E)}\,a_{-1}}
	{E-(E_R-i\G/2)}\\ 	\label{eq:regular}\label{paul19}\label{bohm19}
	&=&-2\pi ia_{-1}\langle\psi^-|z^-_R\rangle 
	e^{2i\gamma(z_R)}\langle^+z_R|\phi^+\rangle\\
	\vspace{.3cm}&=&-e^{2i\gamma(z_R)}2\pi\Gamma\;\langle\psi^-|
	z^-_R\rangle\langle^+z_R|\phi^+\rangle\;.\nonumber 
\end{eqnarray}
The integral in~(\ref{bohm19}) is obtained from the integral in
(\ref{bohm18}) by deforming the contour of integration into the real
axis of the second sheet plus the infinite semicircle and 
omitting in~(\ref{bohm18}) the integral over the infinite semicircle
in the lower half-plane of the second sheet, because it is zero.
Eq.~(\ref{eq:regular}) 
is a special case of the Titchmarsh theorem. The value at $z=z_R$ of
the analytic function $\langle\psi^-|z^-\rangle\,e^{2i\gamma(z)}$
defines a continuous antilinear functional $F(\psi^-)\equiv\langle\psi^-|
z^-_R\rangle\,e^{2i\gamma(z_R)}=\langle\psi^-|z^\gamma_R\rangle$ over the
space $\Phi_+\ni\psi^-$, and this functional establishes the
generalized vector
$|z^\gamma_R\rangle=|z^-_R\rangle\,e^{2i\gamma(z_R)}\in\Phi^\times_+$.

We can rewrite (\ref{eq:regular}) by omitting the arbitrary 
$\psi^-\in\F_+$ and write it as an equation for the functional
$|z^-_R\rangle \in\FT_+$, 
\begin{eqnarray}\label{eq:5.32}\label{paul20}\label{bohm20}
	|z^-_R\rangle &=& \frac{i}{2\pi}\int^{+\infty}_{-\infty_{\rm II}} \, 
	dE\;|E^-\rangle\;\frac{\<^+E|\phi^+\rangle}{\<^+z_R|\phi^+\rangle }\;
	\frac{1}{E-(E_R-i\,\G /2)}\\ \nonumber 
	&=& -\frac{1}{2\pi i}\int^{+\infty}_{-\infty_{\rm II}}
	\,dE\;|E^-\rangle\;\frac{1}{E-z_R}
\end{eqnarray}
over all $\psi^-\in\F_+$. Or we can rewrite~(\ref{bohm19}) as an 
equation for the operator from $\F_-$ (preparation) to $\F_+$
(registration) by omitting the arbitrary $\psi^-\in\Phi_+$ and 
the arbitrary $\phi^+ \in\F_-$:
\begin{equation}\label{eq:5.33}
	|z^-_R\rangle\<^+z_R|=\frac{i}{2\pi}
	\int^{+\infty}_{-\infty_{\rm II}} dE\;
	{{|E^-\rangle\<^+E|}\over {E-(E_R -i\,\G /2)}}\;.
\end{equation}
The notation for the vectors $|z^-_R\rangle$ derives from the 
Cauchy theorem: $\<\psi^-|z^-_R\rangle$ is the value of the function
$\<\psi^-|\om^-\rangle$ at the position $\om=z_R$.  The definition
(\ref{eq:5.32}) of the Gamow vector is, like (\ref{eq:regular}) and
(\ref{eq:5.33}), just another example of the Titchmarsh theorem.  From
the above derivation,  one can see why we defined the Gamow vectors:  They 
are the vectors associated with the pole term of the S-matrix element. The 
``normalization'' of the vectors $|z^-_R\rangle$ is a consequence of
the ``normalization'' of the Dirac kets $|E^-\rangle$, and we can
define Gamow vectors $\psi^G$ with arbitrary normalization and phase $N(z_R)$, 
\begin{subeqnarray}\label{eq:5.34}\label{bohm22}
	\psi^G = |z^-_R\rangle\;N(z_R)\;.\hspace{1.9cm}
\end{subeqnarray}
A normalization that we shall use here is
\begin{equation}\nonumber
	\hspace{5.3cm}\psi^G\equiv |z^-_R\rangle\Bigl(-e^{2i\gamma(z_R)}\Bigr)
	\sqrt{2\pi\Gamma}\;.\hspace{5.2cm}(\ref{bohm22}{\rm b})
\end{equation}
The constant phase factor $-e^{2i\gamma(z_R)}$, which we introduced
in~(\ref{bohm22}b) is arbitrary and of no significance here, and the
``normalization'' factor $\sqrt{2\pi\Gamma}$ is also a matter of
convention.(\cite{bohm}, sect.~XXI.4) 

The notation $|z^-_R\rangle$ has a further meaning: It can be
shown~\cite{bohm,bohm-gadella} that this vector is a generalized
eigenvector of the self-adjoint~\cite{22a} Hamiltonian $H$ with eigenvalue
$z_R=E_R-i\Gamma/2$: 
\begin{equation}\label{bohm23}
	\<\psi^-|H^\times \psi^G\rangle\equiv\< H\psi^-|\psi^G\rangle =
	z_R\;\<\psi^-|\psi^G\rangle\,,\qquad\forall\, 
	\psi^- \hspace{-.05in} \in\F_+\;.
\end{equation}
where $H^\times$ is the conjugate operator in $\Phi^\times$ of the
operator $H$ in $\Phi$. This one writes as
\begin{equation}\label{bohm24}
	H^\times\psi^G=z_R\;\psi^G\hspace{.3in}\qquad\text{or~also}\qquad
	H^\times|z^-_R\rangle=z_R\,|z^-_R\rangle
\end{equation}
or following Dirac's notation $H|E^-\rangle=E|E^-\rangle$ 
\begin{equation}\nonumber
	H\psi^G=z_R\;\psi^G\hspace{.3in}\qquad\text{or~also}\qquad
	H|z^-_R\rangle=z_R\,|z^-_R\rangle
\end{equation}
if the operator $H$ is essentially self-adjoint. If one takes the
complex conjugate of (\ref{bohm23}) one obtains: 
\begin{equation}\label{eq:5.34prime}
	\<\psi^G |\, H\, |\psi^-\rangle =  \<\psi^G|\psi^-\rangle 
	\Bigl( E_R +i\, \frac{\G}{2}\Bigr)
\end{equation}
which one can write in analogy to~(\ref{bohm24}) as
\begin{equation}\label{bohm26}
	\langle \psi^G |\, H =  z^\ast_R\, \langle \psi^G |\, \hspace{.3in}
	\text{or}\hspace{.3in} 
	\langle^-z_R |\,   H =  z^\ast_R\, \langle^- z_R  | \;.
\end{equation}

It has also been shown~\cite{bohm,bohm-gadella} that in the 
RHS~(\ref{bohm7}b) the time evolution is given by a semigroup operator
\begin{equation}\label{eq:5.88}
	U^\times_+ (t)\equiv U(t)|^\times_{\F_+}\equiv\left(\left.
	e^{iHt}\right|_{\F_+}\right)^\times\equiv e^{-iH^\times t}_+
	\;;\hspace{.25in}{\rm for~}t\ge0 
\end{equation}
(A similar semigroup time evolution operator $e^{-iH^\times t}_-$,
defined however only for $t\le 0$, also exists in the RHS
(\ref{eq:rhsm}a) and has similar properties.)
And it has been shown that this time evolution operator
(\ref{eq:5.88}) acts on the Gamow vectors
$\psi^G$ (or on the $|z^-_R\rangle\in\FT_+$) in the following way:
\begin{equation}\label{bohm28}
	\<\psi^-|e^{-iH^\times t}_+|z^-_R\rangle \equiv
	\<e^{iHt}\psi^-|z^-_R\rangle
   	=e^{-iE_R t}e^{-(\G / 2)t}\, \<\psi^-|z^-_R\rangle
\end{equation}     
or for the complex conjugate
\begin{equation}\label{bohm29}
	\hspace{-.5cm}\langle^-z_R|\,e^{iHt}\,|\psi^-\rangle=e^{iE_Rt}e^{-(\G 
	/2)t}\langle^-z_R|\psi^-\rangle\hspace{.25in}
	\begin{array}{l}
		{\rm for~every~}\psi^-\in\F_+\\{\rm and~for~}t\ge0\;.
	\end{array}
\end{equation}
Omitting the arbitrary $\psi^-\in\F_+$, this is also written in
analogy to~(\ref{bohm24}) as
\begin{subeqnarray}\label{eq:5.89prime}\label{paul30}\label{bohm30}
	e^{-iH^\times t}_+\,\psi^G&=&e^{-iE_R t}e^{-(\G /2)t}\,\psi^G\\
	{\rm or}\hspace{5.1cm}&&\hspace{5.4cm}\hspace{3.8cm}\nonumber\\
	\langle\psi^G |\,e^{iHt}&=&e^{+iE_R t}e^{-(\G /2)t}\,
	\langle\psi^G|\hspace{.5in}{\rm for~}t\ge 0\;.
\end{subeqnarray}

One of the most important features of the Gamow vectors is that they
are basis vectors of a basis system expansion.  To explain this we
start with the Dirac basis vector expansion (the Nuclear Spectral
Theorem of the rigged Hilbert space) which states that
\begin{equation}\label{bohm31}
	\hspace{-.15in}\phi = \int^{+\infty}_0\,dE\;|E^+\rangle
	\langle^+ E|\phi^+\rangle\;+\;\sum_m\;|E_m)(E_m|\phi)
	\hspace{.25in}{\rm for~every~}\phi\in\F\;.
\end{equation}
In here, $|E_m)$ are the discrete eigenvectors of the exact
Hamiltonian $H=K+V$, describing the bound states, $H\, |E_m)= E_m\,
|E_m)$, and $|E^+\rangle$ are the generalized eigenvectors (Dirac
kets) of $H$, describing scattering states~\cite{fussnote}. The 
integration extends over the continuous spectrum of $H: \; 0\le E <\infty$.

Instead of the basis vector expansion~(\ref{bohm31}) which uses Dirac
kets that correspond to the (continuous) spectrum of
$H$, one can use a basis system that contains Gamow vectors,
and one obtains the so-called ``complex basis vector expansion'' which states:
For every $\phi^+\in\F_-$ (a similar expansion holds also for every 
$\psi^-\in\F_+$), one obtains for the case of a finite number of first
order (resonance) poles at the positions $z_{R_i}$, $i=1,\, 2,\, \dots
,N$, the following basis system expansion:
\begin{eqnarray}\label{eq:basisexpansion}\label{paul32}\label{bohm32}
	\phi^+&=&\int^{-\infty_{\rm II}}_0\,dE|E^+\rangle
	\langle^+E|\phi^+\rangle\;-\;\sum^N_{i=1}|z^-_{R_i}\rangle
	2\pi\G_i\,e^{2i\gamma(z_{R_i})}\langle^+ z_{R_i}|\phi^+\rangle\\ 
	&&+\sum_m|E_m)(E_m|\phi^+)\hspace{1in}\text{for }\phi^+\in\F_-\nonumber
\end{eqnarray}
where $-|z^-_{R_i}\rangle\sqrt{2\pi\Gamma}\,e^{2i\gamma(z_R)}=\psi^{G_i}
\in\F^\times_+$ are Gamow vectors representing decaying states.  The
first integral in~(\ref{paul32}) comes from the ``background term'' of
equation (\ref{bohm17}). This background integral is omitted in the
phenomenological theories with a complex effective
Hamiltonian~\cite{lee, lee-oehme-yang}. The integration
in~(\ref{paul32}) is along the negative real axis in the second sheet
or along an equivalent contour. The third term will be absent if there is 
no bound state $|E_m)$, which we shall assume from now on.
The expansion (\ref{eq:basisexpansion}) follows directly from
(\ref{bohm17}) for $r=1$ (if one assumes that the S-matrix has no
other singularities in the lower half-plane besides the $N$ first order
poles at the positions $z_{R_i}$, which is a realistic assumption if
one excludes higher order poles~\cite{nuss}).

The matrix representation of $H$ in the basis of (\ref{bohm31})
is given by:
\begin{align}\label{eq:5.61a}
	\left( \begin{matrix} \<\psi |\, H^\times \, |E_1) \\
                 \<\psi |\, H^\times \, |E_2) \\
                      \,\vdots \\
                 \<\psi |\, H^\times \, |E_N) \\
 	         \< \psi |\, H^\times \, |E^\pm\rangle 
  	\end{matrix} \right) 
      =	\begin{pmatrix} E_1 & 0 & \cdots &\cdots& 0   \\
                      0 & E_2 &&& 0          \\
                      \vdots && \ddots && \vdots  \\
                      \vdots &&& E_N & 0     \\
                       0 & 0 & \cdots & 0 & (E)           
	\end{pmatrix}
     	\begin{pmatrix}  \<\psi |E_1 ) \\
                      \<\psi |E_2 ) \\
                       \,\vdots     \\
                      \<\psi |E_N ) \\
                      \<\psi |E^\pm\rangle 
	\end{pmatrix}						\\\nonumber
	0\le E <+\infty 
\end{align}
for all $\psi\in\F^{\times\times}=\F$.  In (\ref{eq:5.61a}), 
the operator $H^\times$ is represented by a finite or an infinite
diagonal submatrix (for a finite or an infinite number of bound
states) and a continuously infinite diagonal submatrix, indicated 
by $(E)$, where $E$ takes the values $0\le E<+\infty $.  If we
consider only the case where there are no bound states (meaning we
omit the submatrix of the $E_m$) then the matrix representation
corresponding to the basis system expansion~(\ref{bohm31})  
is simply given by the diagonal continuously infinite
real energy matrix:
\begin{equation}\label{eq:5.61b}
	\Bigl(\<H\psi|E^\pm\rangle\Bigr) = 
	\Bigl(\<\psi|\,H^\times\,|E^\pm\rangle\Bigr) =
	\Bigl(E\Bigr)\Bigl(\<\psi|E^\pm\rangle\Bigr)\,;\quad
	\psi\in\F,\quad 0\le E<+\infty
\end{equation}

On the other hand, the complex basis vector expansion
(\ref{eq:basisexpansion}) (again without bound states) leads to 
a matrix representation of the self-adjoint semibounded 
Hamiltonian $H$ in the following form:
\begin{align}\label{eq:5.62}
	\hspace{-0.4in}
	\left(\begin{matrix} \<H\psi^-|z^-_{R_1}\rangle \\
                \<H\psi^-|z^-_{R_2}\rangle \\ 
                    \vdots \\
                \<H\psi^-|z^-_{R_N}\rangle \\ 
           	\<H\psi^-|E^-\rangle  
  	\end{matrix} \right)
      =	\left(\begin{matrix} \<\psi^-|\, H^\times \, |z^-_{R_1}\rangle \\ 
                \<\psi^-|\, H^\times \, |z^-_{R_2}\rangle \\ 
                    \vdots \\
                \<\psi^-|\, H^\times \, |z^-_{R_N}\rangle \\ 
           	\<\psi^-|\, H^\times \, |E^-\rangle  
  	\end{matrix} \right) 
      =	\begin{pmatrix} z_{R_1} &&&& 0    \\
                      & z_{R_2 } &&& 0  \\
                      && \ddots & & \vdots   \\
                      &&& z_{R_{N}} & 0 \\
                      0 & 0 & \ldots & 0 & (E) 
	\end{pmatrix}
      	\begin{pmatrix} \<\psi^- |z^-_{R_1} \rangle \\
                      \<\psi^- |z^-_{R_2} \rangle \\
                       \,\vdots                   \\
                      \<\psi^- |z^-_{R_N}\rangle   \\
                      \<\psi^- |E^-\rangle   
	\end{pmatrix} 						\\\nonumber
	\psi^-\in\F_+\subset\F \quad -\infty_{\rm II} < E\le 0  
\end{align}
The same Hamiltonian $H$ with $N$ resonances at $z_{R_i}$, $i=1,\,
2,\, \ldots N$, can thus be 
represented either as a continuous infinite matrix (\ref{eq:5.61b}) 
in the basis of (\ref{bohm31}), or by (\ref{eq:5.62}) in the
basis of (\ref{eq:basisexpansion}).  The later alternative is of 
more practical importance if one wants to study the resonance
properties and if one can make $\<\psi^-|E^-\rangle$ small.  
The basis vector expansion~(\ref{paul32}) is an exact representation
of $\phi^+\in\Phi_-$ and the matrix representation (\ref{eq:5.62}) 
is an exact representation of the self-adjoint Hamiltonian.  In the
phenomenological descriptions by complex effective Hamiltonians, one
uses a truncation of~(\ref{paul32}) and~(\ref{eq:5.62}), omitting the
background integral in~(\ref{paul32}) and 
the whole continuously infinite diagonal matrix $\bigl(E\bigr)$ (and
sometimes even some of the $z_{R_i}$) in~(\ref{eq:5.62}). In this
approximation one represents the Hamiltonian
by the $N\times N$ dimensional diagonal complex submatrix in the
upper left corner of (\ref{eq:5.62}). For example, if one considers
only two resonances at $z_{R_1}=z_S \,, \;z_{R_2}=z_L$, one then has
the complex energy matrix: 
\begin{equation}\label{eq:5.63}
	\left( \begin{matrix} \<\psi^-|\, H^\times\, |z^-_S\rangle \\
              	\<\psi^-|\, H^\times \, |z^-_L\rangle 
	\end{matrix} \right) 
      =	\begin{pmatrix} z_S & 0 \\
                0 & z_L 
	\end{pmatrix}
    	\begin{pmatrix} \<\psi^-|z^-_S\rangle \\
             	\<\psi^-|z^-_L\rangle 
	\end{pmatrix}
\end{equation}
This truncated matrix representation is only an approximation, 
corresponding to the approximation of omitting the integral in 
(\ref{eq:basisexpansion}).  How good this approximation is depends upon 
the particular choice of the $\psi^-$ (or the choice of the 
$\phi^+$), but it can never be exact.

\section{Higher Order Poles of the S-matrix and Gamow-Jordan
Vectors} \label{sec:GJvectors}\label{sec:higher} 
\setcounter{equation}{0}

We shall now discuss the possibility of extending the definition of
one generalized eigenvector $|z^-_R\rangle^{(0)}$ to $r$ generalized
eigenvectors of order {${n=0,~1,~2,\dots,r\!-\!1}$} for an S-matrix pole of
order $r$.~\cite{eigenvector} ~ 
The equations~(\ref{bohm18}) and~(\ref{bohmprime}) for the pole 
term are rewritten (omitting on the right-hand side the integral over
the infinite semicircle in the lower half-plane of the second sheet) as
\begin{eqnarray}
	\frac{i}{2\pi}(\psi^-,\phi^+)_{\rm P.T.}&=&\sum_{n=0}^{r-1}
	\frac{i}{2\pi}\int^{+\infty}_{-\infty_{\rm II}}dE\,
	\langle\psi^-|E^-\rangle\frac{e^{2i\gamma(E)}\,a_{-n-1}}
	{(E-z_R)^{n+1}}\langle^+ E|\phi^+\rangle\nonumber\\ 
	&=&\sum_{n=0}^{r-1}\frac{1}{n!}\,a_{-n-1}\,\frac{d^n\;}{d\om^n} 
   	\Bigl( \langle\psi^- |\om^\gamma\rangle\langle^+ \om|\phi^+\rangle
   	\Bigr)_{\om =z_R}			\label{eq:41}\label{bohm37}\\
	&=&\sum_{n=0}^{r-1}\frac{1}{n!}\,a_{-n-1}\sum^n_{k=0}
	\begin{pmatrix}n\\k\end{pmatrix}\langle\psi^- |z^\gamma_R \rangle^{(k)}
	\,^{(n-k)}\langle^+z_R|\phi^+\rangle\nonumber\\
	&=&\sum_{n=0}^{r-1}\frac{i}{2\pi n!}\int^{+\infty}_{-\infty_{\rm II}}dE
	\frac{\left(\langle\psi^-|E^-\rangle e^{2i\gamma(E)}a_{-n-1}
	\langle^+E|\phi^+\rangle\right)^{(n)}}{E-z_R}\nonumber
\end{eqnarray}
Since $G_-(E)=\langle\psi^-|E^-\rangle\langle^+E|\phi^+\rangle\in{\cal
S}\cap{\cal H}_-^1$, its $(n+1)$-st order derivatives are also elements
of ${\cal S}\cap{\cal H}^1_-$, and~(\ref{bohm37}) is an application of
the Titchmarsh theorem in two different versions, for
$G_-(E)=\langle\psi^-|E^-\rangle\langle^+E|\phi^+\rangle$ and for
$G_-(E)=(\langle\psi^-|E^-\rangle\langle^+E|\phi^+\rangle)^{(n)}$.

The value at $z=z_R$ of the analytic functions
$\langle\psi^-|z^\gamma\rangle^{(k)}$ ($k$-th derivatives of the
analytic function $\langle\psi^-|z^\gamma\rangle$) defines again a
continuous antilinear functional
$F^k(\psi^-)\equiv\langle\psi^-|z^\gamma_R\rangle^{(k)}$ over the space
$\Phi_+\ni\psi^-$. The antilinearity follows from the linearity of the
differentiation 
$\left(\langle\alpha\psi^-_1+\beta\psi^-_2|z\rangle\right)^{(k)} 
=\alpha^*\langle\psi_1^-|z\rangle^{(k)}+\beta^*\langle
\psi^-_2|z\rangle^{(k)}$. The continuity follows because 
taking the $k$-th derivative $D^k$ is a continuous
operation with respect to the topology in the space ${\cal S}\cap{\cal
H}^2_-\ni\langle\psi^-|E^\gamma\rangle$ and because 
$\langle\psi^-|z^\gamma_R\rangle^{(k)}$ is a continuous functional
$F$. Thus $F^k\equiv D^k\circ F$ is the product of two continuous maps and
therefore also continuous. The continuous functionals
$\langle\psi^-|z^\gamma_R\rangle^{(k)}$ define thus the generalized
vectors $|z^\gamma_R\rangle^{(k)}\in\Phi^\times_+$, $k=0,~1,\dots,~r-1.$
The $r$-th order pole is therefore by~(\ref{bohm37}) associated 
with the set of $r$ generalized vectors 
\begin{equation}\label{bohm38}
	|z^\gamma_R\rangle^{(0)}\,,\;|z^\gamma_R\rangle^{(1)}\,,\;\cdots,
	|z^\gamma_R\rangle^{(k)}\,,\cdots,|z^\gamma_R\rangle^{(n)}\;.
\end{equation}
Of the different representations of the pole term on the right-hand
side of~(\ref{bohm37}) we shall use in this paper only the second and
third line and will come back to the integral representations when we
discuss the Golden Rule for the higher order Gamow states.

We insert the values~(\ref{b17b}) of the coefficients $a_{-n-1}$
into~(\ref{eq:41}) and obtain
\begin{eqnarray}\label{bohm39}
	(\psi^-,\phi^+)_{\rm P.T.}
	&=&\;-\sum_{n=0}^{r-1}\begin{pmatrix}\!r\!\\\!n\!+\!1\!\end{pmatrix}
	\frac{(-i)^n}{n!}\frac{d^n}{d(\omega/\Gamma)^n}
	\Bigl(\langle\psi^-|\omega^\gamma\rangle\,2\pi\Gamma\,
	\langle^+\omega| \phi^+\rangle\Bigr)_{\omega=z_R}\\
	&=&\;-\sum_{n=0}^{r-1}\begin{pmatrix}\!r\!\\
	\!n\!+\!1\!\end{pmatrix}\frac{(-i\Gamma)^n}{n!}2\pi\Gamma\sum^n_{k=0}
	\begin{pmatrix}n\\k\end{pmatrix}\langle\psi^-|z^\gamma_R\rangle^{(k)}
	\,^{(n-k)}\langle^+ z_R | \phi^+\rangle
	\nonumber
\end{eqnarray}
The generalized vectors~(\ref{bohm38}) have all different dimensions,
namely $({\rm energy})^{-\frac12-k}$. If one uses the dimensionless
variable $\omega/\Gamma$ as indicated in the first line 
of~(\ref{bohm39}), one is led to the new normalization of the
generalized vectors
\begin{equation}\label{bohm40}
	|z^\gamma_R\mer^{(k)}=\frac{1}{k!}|z^\gamma_R\rangle^{(k)}
	\;\Gamma^k\hspace{.3in}{\rm and}\hspace{.3in}
	\,^{(l)}\klam^+z_R|=\Gamma^l\;^{(l)}\!\langle^+z_R|\frac{1}{l!}
\end{equation}
These vectors have for all values of $k=0,~1,~2,\dots,r\!-\!1\;$ the
same dimension  $({\rm energy})^{-\frac12}$, like the Dirac kets. We
have in addition introduced the factor $1/k!\;$ so that these higher
order Gamow vectors become Jordan vectors with the standard
normalization. 
The quantity $\langle\psi^-|z^-\mer^{(n)}\equiv\frac{\Gamma^n}{n!}
\langle\psi^-|z^-\rangle^{(n)}$ is the value of the functional
$|z^-\mer^{(n)}\in\Phi^\times_+$ at $\psi^-\in\Phi_+$. However, unlike
$\langle\psi^-|z^-\rangle^{(n)}$, which is the $n$-th
derivative of $\langle\psi^-|z^-\rangle\in{\cal S}\cap{\cal H}^2_+$,
the $\langle\psi^-|z^-\mer^{(n)}$ is not the $n$-th derivative of 
$\langle\psi^-|z^-\mer^{(0)}$;
the standard Jordan vectors $|z^-\mer^{(k)}$ are
connected with the ``derivatives'' $|z^-\rangle^{(k)}$ by~(\ref{bohm40}).
Therefore when we want to compare our results with the standard
results in the theory of finite dimensional complex
(non-diagonalizable) matrices~\cite{baumgartel-gantmacher-lancaster}
we need to convert from the $|z^-\rangle^{(k)}$ to the $|z^-_R\mer^{(k)}$.

With the convention~(\ref{bohm40}) we obtain from~(\ref{bohm39}) 
\begin{eqnarray}\label{bohm41}
	(\psi^-,\phi^+)_{\rm P.T.}
	&=&-\sum_{n=0}^{r-1}\begin{pmatrix}r\\n+1\end{pmatrix}
	(-i)^n(2\pi\Gamma)\sum^n_{k=0}\langle\psi^-|z^\gamma_R\mer^{(k)}
	\,^{(n-k)}\klam^+ z_R | \phi^+\rangle\\
	&=&-2\pi\Gamma\,\sum_{n=0}^{r-1}\begin{pmatrix}r\\n+1\end{pmatrix}
	(-i)^n\langle\psi^-|{W^\gamma}^{(n)}|\phi^+\rangle\nonumber	
\end{eqnarray}
where we have defined the operator
\begin{equation}\label{bohm42}
	\hspace{-.2in}W_{\rm P.T.}^{(n)}=\sum_{k=0}^n|z^-_R\mer^{(k)}
	\,^{(n-k)}\klam^+z_R|\hspace{.2in}{\rm and}\hspace{.2in}
	W^{\gamma(n)}_{\rm P.T.}=\sum_{k=0}^n|z^\gamma_R\mer^{(k)}
	\,^{(n-k)}\klam^+z_R|\;.
\end{equation}
Here $W^{\gamma(n)}_{\rm P.T.}$ is just an abbreviation for the
right-hand side of~(\ref{bohm42}), and in section~\ref{sec:possible}
we will discuss its interpretation. Whereas $W^{\gamma(n)}_{\rm
P.T.}$ depends also upon the background phase shifts
through~(\ref{b18a}), $W_{\rm P.T.}^{(n)}$ is just given by the
S-matrix pole.

We now return to the complete S-matrix element~(\ref{bohm17}) and
insert the pole term~(\ref{bohm41}) into~(\ref{bohm17}b),
\begin{eqnarray}\label{bohm43}
	(\psi^-,\phi^+)&=&\int_0^{-\infty_{\rm
	II}}dE\,\langle\psi^-|E^-\rangle\,  
	S_{\rm II}(E)\,\langle^+E|\phi^+\rangle\\\nonumber
	&&\!\!\!-\sum_{n=0}^{r-1}
	\begin{pmatrix}r\\\!n\!+\!1\!\end{pmatrix}(-i)^n2\pi\Gamma
	\sum_{k=0}^n\langle\psi^-|z^\gamma_R\mer^{(k)}
	\;^{(n-k)}\klam^+\!z_R|\phi^+\rangle			
\end{eqnarray} 
Omitting the arbitrary $\psi^-\in\Phi_+$ and
rearranging the sums in the second term, we obtain the complex basis 
vector expansion for an arbitrary $\phi^+\in\Phi_-$,
\begin{eqnarray}\label{bohm44}\label{eq:50}
	\phi^+&=&\int_0^{-\infty_{\rm II}}dE|E^+\rangle\langle^+E|
	\phi^+\rangle\,+\\
	&&+\,\sum_{k=0}^{r-1}|z^\gamma_R\mer^{(k)}
	\left((-2\pi\Gamma)\sum_{n=k}^{r-1}
	\begin{pmatrix}r\\n+1\end{pmatrix}(-i)^n
	\,^{(n-k)}\klam^+z_R|\phi^+\rangle\right)\nonumber		
\end{eqnarray}
This complex basis vector expansion is the analogue of~(\ref{bohm32})
if instead of $N$ S-matrix poles of order one (and bound states $|E_m)$)
we have one S-matrix pole of order $r$. To compare~(\ref{bohm44})
with~(\ref{bohm32}), we write~(\ref{bohm32}) also for the
case of one S-matrix pole of order one. Then using the same
phases~(\ref{bohm19}) as in~(\ref{bohm39}) and omitting all bound
states and all resonances but one, we obtain for~(\ref{bohm32})
\begin{equation}\label{bohm45}
	\phi^+=\int_0^{-\infty_{\rm II}}dE|E^+\rangle\langle^+E|\phi^+
	\rangle-|z^\gamma_R\rangle\;2\pi\Gamma\langle^+z_R|\phi^+\rangle
\end{equation}
which agrees with what we obtain from~(\ref{bohm44}) for
$r=1$. Comparing~(\ref{bohm44}) with~(\ref{bohm45}) or~(\ref{bohm32}) we see
the similarities and the differences: For a first order pole there is
one generalized vector in the complex basis vector expansion; for an
$r$-th order pole there are $r$ basis vectors in the complex basis
vector expansion. Apart from the arbitrary phase-normalization factor
$-2\pi\Gamma$, the coefficient of the first order Gamow vector,
$|z^\gamma_R\rangle=|z^-_R\rangle\,e^{2i\gamma(z_R)}$, has the
simple form $\langle^+z_R|\phi^+\rangle$ which resembles the 
component $\langle^+E|\phi^+\rangle$ of the vector $\phi^+$ along the
basis vector $|E^+\rangle$. In contrast, the coefficients of the
higher order Gamow vectors $|z^\gamma_R\mer^{(k)}$ are 
given by the complicated expression 
\begin{equation}\label{bohm46}
	b_k=(-2\pi\Gamma)\sum_{n=k}^{r-1}\begin{pmatrix}r\\n+1
	\end{pmatrix}(-i)^n\,^{(n-k)}\klam^+z_R|\phi^+\rangle\;.
\end{equation}
The difference between~(\ref{bohm44}) and~(\ref{bohm32}) also
foretells that the role of dyadic products like
$\left.|E_n\right)\left(E_n|\right.$ (or also
of $|z^-_R\rangle\langle^-z_R|$), which have been prominently used for
pure states, will probably be unimportant for states associated with
higher order poles. In section~\ref{sec:possible}, we will see that for
higher order Gamow states there is no meaning to being pure.

Since the general expressions~(\ref{bohm44}) and~(\ref{bohm43}) are not
very transparent, we want to specialize them now to the case of a
double pole, $r=2$:
\begin{subeqnarray}\nonumber
	\phi^+&=&\int_0^{-\infty_{\rm II}}dE|E^+\rangle
	\langle^+E|\phi^+\rangle +\\\label{bohm47}\label{4.11a}
	&&-|z^\gamma_R\mer^{(0)}2\pi\Gamma
	\Bigl(2\;^{(0)}\klam^+z_R|\phi^+\rangle-i	
	\;\;^{(1)}\klam^+z_R|\phi^+\rangle\Bigr)\\
 	&&+|z^\gamma_R\mer^{(1)}2\pi\Gamma
	i\;\;^{(0)}\klam^+z_R|\phi^+\rangle\nonumber
\end{subeqnarray}
If as a generalization of~(\ref{bohm22}b), we define the differently
normalized Gamow vectors
\begin{subeqnarray}\label{4.12a}
	\psi^{G(k)}=(-1)^{k+1}|z^\gamma_R\mer\!^{(k)}\sqrt{2\pi\Gamma}
	=(-1)^{k+1}\frac{\Gamma^k}{k!}|z_R^\gamma\rangle^{(k)}\sqrt{2\pi\Gamma}
\end{subeqnarray}
then the basis vector expansion~(\ref{bohm47}a) for the case $r=2$ reads
\begin{eqnarray}\nonumber
	\hspace{3.2cm}\phi^+&=&\int_0^{-\infty_{\rm II}}dE|E^+\rangle
	\langle^+E|\phi^+\rangle+\\	
	&&+\psi^{G(0)}\sqrt{2\pi\Gamma}\Bigl(-2\;^{(0)}\!\langle^+z_R
	|\phi^+\rangle+(-i)\;^{(1)}\!\langle^+z_R|\phi^+\rangle\Bigr)
	\hspace{1.8cm}(\ref{bohm47}{\rm b})\nonumber\\
	&&+\psi^{G(1)}\sqrt{2\pi\Gamma}\;i\;^{(0)}\!\langle^+z_R
	|\phi^+\rangle\;.\nonumber
\end{eqnarray}
Note that according to~(\ref{4.12a}a) and~(\ref{b18a}) we have 
\begin{equation}\nonumber
	\hspace{3.2cm}\psi^{G(1)}=\Gamma\Bigl(|z^-_R\rangle^{(1)}
	+|z^-_R\rangle 2i\gamma'(z_R)\Bigr)e^{2i\gamma(z_R)}
	\sqrt{2\pi\Gamma}\hspace{3.8cm}(\ref{4.12a}{\rm b})
\end{equation}
and only for constant background phase shift $\gamma^{(n)}(z)=0$,
$n=~1,~2,\dots$, $\psi^{G(1)}$ (or $\psi^{G(k)}$) given by
$|z^-_R\rangle^{(0)}$ (or $|z^-_R\rangle^{(k)}$). One can
insert~(\ref{4.12a}b) into~(\ref{bohm47}b) and expand
$\phi^+$ in terms of the basis vectors $|z^-_R\rangle$ and
$|z^-_R\rangle^{(1)}$; and the same procedure one can repeat for arbitrary
$k$ to express $\phi^+$ in~(\ref{bohm44}) in terms of 
\begin{equation}\label{4.13}
	|z^-_R\rangle^{(0)}\,,\;|z^-_R\rangle^{(1)}\,,\;\cdots,
	|z^-_R\rangle^{(k)}\,,\cdots,|z^-_R\rangle^{(n)}\;.
\end{equation}
Whether the phase convention in the definition~(\ref{4.12a}) will
turn out to be convenient cannot be said at this stage.

The basis vector expansion can be generalized in a straightforward way
to the case of an arbitrary finite number of poles at the positions
{${z_{R_i}, ~~i=1,2,\dots,N}$} of arbitrary finite order $r_i$. 
in the same way as it was done in~(\ref{bohm32}) for $r_i=1$. This complex
generalized basis vector expansion is the most important result of our
irreversible quantum theory (as is the Dirac basis vector expansion
for reversible quantum mechanics).
\noindent It shows that the generalized vectors~(\ref{bohm38})
(functionals over the space $\Phi_+$) are part of a basis system for
the $\phi^+\in\Phi_-$ and form together with the kets
$|E^+\rangle,~~-\infty_{\rm II}<E\leq0$, a complete basis system. The 
vectors~(\ref{bohm38}) span a linear subspace ${\cal
M}_{z_R}\subset\Phi^\times_+$ of dimension $r$:
\begin{equation}\label{bohm48}
	\CM_{z_R} = \Biggl\{ \; \xi \; {\Bigg|} \; 
  	\xi =\sum^{r-1}_{k=0}|z^-_R \rangle^{(k)} c_k\, , \; c_k\in\C\, 
	\Biggr\} \subset\F^\times_+ 
\end{equation}
If there are $N$ poles at $z_{R_i}$ of order $r_i$, then for every
pole there is a linear subspace ${\cal M}_{z_{R_i}}\subset\Phi^\times_+$.
Since the generalization to $N$ poles of order $r_i$ at energy
$z_{R_i}$ is straightforward, we continue our discussions for the case
of one pole of order $r$. 

Note that by the procedure described in this section a new label $k$
was introduced for the basis vectors in the expansion~(\ref{bohm44}),
$|z^-_R\mer^{(k)}=|z_R,b_2,b_3,\dots,b_{\cal N}^-\mer^{(k)}$.
Usually basis vector labels are quantum numbers associated with eigenvalues
of a complete system of commuting observables. That means that if in
addition to $H$ there are the ${\cal N}-1$ operators 
$B_2,B_3,\dots,B_{\cal N}$ with eigenvalues $\{b_2,b_3,\dots,b_{\cal
N}\}\equiv\{b\}={\rm spectrum}(B_2,B_3,\dots,B_{\cal N})$, then the
Dirac kets are labelled by $|E,b^-\rangle=|E,b_2,b_3,\dots,b_{\cal N}^-
\rangle$ and in addition to the sum and integral in~(\ref{bohm31})
and~(\ref{bohm44}), there is a sum and/or an integral over all the
values of the degeneracy quantum numbers $b_2,b_3,\dots,b_{\cal N}$,
which we suppress here for the sake of simplicity. The label $k$ of
the higher order Gamow vectors $|z_R,b^-\rangle^{(k)}$, which has
appeared in~(\ref{bohm44}), is not associated with a conventional quantum
number and is there in addition to the labels $b$ connected with
the eigenvalues of the set of commuting observables
$B_2,B_3,\dots,B_{\cal N}$. The quantum numbers 
$z_R,b_2,b_3,\dots,b_{\cal N}$ can be observed and have an
experimentally defined physical meaning. It is not clear that the
label $k$ will have a similar physical interpretation.
This means that (if a higher order S-matrix pole has at all a physical
meaning) the different vectors $|z^-_R\rangle^{(k)}$ in the subspace
${\cal M}_{z_R}$ have no separate physical meaning (unless $k$ can be
given a physical interpretation).

Now that~(\ref{bohm44}) has established the generalized vectors~(\ref{bohm38})
or the generalized vectors~(\ref{4.13})
as members of a basis system (together with the
$|E^+\rangle;~~0\geq E>-\infty_{\rm II}$) in $\Phi^\times_+$, we can obtain
the action of the operator $H$ by the action of the operator
$H^\times$ on these basis vectors; and we can write the operator $H$ in
terms of its matrix elements with these basis vectors. This can also
be done in the same way for any of the operators $f^*(H)$,
where $f(z)$ is any holomorphic function such that 
\begin{equation}\label{bohm49}
	f^\ast (H) :\F_+\longrightarrow \F_+ \hspace{.3in} 
	\text{is a $\tau_{\F_+}$-continuous operator},
\end{equation} 
(e.g., $f^\ast (H) = e^{iHt}\, ,\; f(H^\times )= e^{-iH^\times t}_+$ for
the real parameter $t\ge 0$ only, since for $t<0$ $f^*(H)=e^{iHt}$ is
not a continuous operator in $\Phi_+$.)
For this purpose we replace the arbitrary $\psi^-\in\Phi_+$
in~(\ref{bohm39}) by $\tilde{\psi}^-=f^*(H)\psi^-$ which is again an
element of $\Phi_+$, because $f^*(H)$ is a continuous operator in
$\Phi_+$ (by assumption~(\ref{bohm49})). Then we obtain by comparing
powers of~$\Gamma$:
\begin{eqnarray}\nonumber
	\sum_{k=0}^n\begin{pmatrix}n\\k\end{pmatrix}\langle f^*(H)
	\psi^-|z^\gamma_R\rangle^{(k)}\;^{(n-k)}\langle^+z_R|
	\phi^+\rangle&=&\frac{d^n}{d\omega^n}\left(f(\omega)
	\langle\psi^-|\omega^-\rangle e^{2i\gamma(\omega)}
	\langle^+\omega|\phi^+\rangle\right)_{\omega=z_R}\\
	&&\hspace{1.75cm}n=0,1,\dots,r-1\;.\label{b50}
\end{eqnarray}
where we have used~(\ref{b18a}) and 
\begin{equation}\label{b51}
	\langle 
	f^*(H)\psi^-|\omega\rangle=\langle\psi^-|f(H^\times)|\omega^-\rangle 
	=f(\omega)\langle\psi^-|\omega^-\rangle 
\end{equation}
which follows from~(\ref{bohm49}). The function 
\begin{equation}\label{b52}
	G(z)\equiv f(z)\langle\psi^-|z^-\rangle\langle^+z|\phi^+
	\rangle e^{2i\gamma(z)}
\end{equation}
is an element of ${\cal S}\cap{\cal H}^2_-$, since 
$\langle\psi^-|z^-\rangle\langle^+z|\phi^+\rangle\in{\cal S}\cap{\cal
H}^2_-$ and $e^{2i\gamma(z)}$ as well as $f(z)$ are holomorphic.
Therefore we can take the derivatives $G(z)^{(n)}$ of any order
\begin{equation}\label{b53}
	G(z)^{(n)}=\sum_{k=0}^n\begin{pmatrix}n\\k\end{pmatrix}
	\left(f(z)\langle\psi^-|z^\gamma\rangle\right)^{(k)}
	\;^{(n-k)}\langle^+z|\phi^+\rangle\;.
\end{equation}
Inserting this into~(\ref{b50}) we obtain:
\begin{eqnarray}\nonumber
	&&\hspace{-.3in}\sum^n_{k=0}\left(\langle\psi^-|f(H)|
	z^\gamma_R\rangle^{(k)}-\left( f(z)\langle\psi^-|
	z^\gamma\rangle\right)^{(k)}_{z=z_R}\right) 
        \begin{pmatrix}n\\k\end{pmatrix}\,^{(n-k)}
	\langle^+ z_R|\phi^+\rangle=0\\				\label{b54}
	&&\hspace{2.8in}n=0,1,\dots,r-1
\end{eqnarray}
Since this has to hold for every $\phi^+\in\F_-$ (i.e., for every
$\langle^+ E|\phi^+\rangle \in\CS\cap\HH^2_-$), it follows that the
coefficients of each derivative $\,^{(n-k)}\langle^+ z_R
|\phi^+ \rangle = \frac{d^{(n-k)}\,}{dz^{(n-k)}} \left. \langle^+ z
|\phi^+\rangle \right|_{z=z_R}$ must vanish. Thus,
\begin{equation}\label{b55}
	\langle\psi^-|\,f(H^\times)\,|z^\gamma_R\rangle^{(k)}
	=\Bigl(f(z)\langle\psi^-|z^\gamma\rangle\Bigr)^{(k)}_{z=z_R}
	\hspace{.3in}\begin{array}{l}\text{for $k=0,\, 1,\, 2,\dots n$}\\
	\text{and all $\psi^-\in\F_+$\;.}\end{array}
\end{equation}
By a similar argument, just comparing the coefficients of 
$(e^{2i\gamma(z)}\langle^+z|\phi^+\rangle)^{(n-k)}_{z=z_R}$ rather
than of $(\langle^+z|\phi^+\rangle)^{(n-k)}_{z=z_R}$, one can show that
the same equation holds for the $|z^-_R\rangle^{(k)}$ (with any nice
function for $\gamma(z)$):
\begin{equation}\label{b55'}
	\langle\psi^-|\,f(H^\times)\,|z^-_R\rangle^{(k)}
	=\Bigl(f(z)\langle\psi^-|z^-\rangle\Bigr)^{(k)}_{z=z_R}
\end{equation}

This permits us to calculate the action of $f(H^\times )$ on the
generalized vectors $|z^-_R\rangle^{(k)}\in\F^\times_+$ for every 
$f(H^\times )$ that fulfills the condition (\ref{bohm49}).  
The same calculation applies to the generalized vectors
$|z^\gamma_R\rangle^{(k)}$ due to~(\ref{b55}). Therefore we write the
following equations for $|z^\gamma_R\rangle^{(k)}$ though the same
holds for $|z^-_R\rangle^{(k)}$.

We first choose $f(H^\times )= H^\times$; then we obtain
\begin{equation}\label{b56}
	\langle H\psi^-|z^\gamma_R\rangle^{(k)}\equiv\langle\psi^-|H^\times 
	|z^\gamma_R\rangle^{(k)}=z_R\langle\psi^-|z^\gamma_R\rangle^{(k)}+ 
	\begin{pmatrix}k\\1\end{pmatrix}\langle\psi^-|z^\gamma_R\rangle^{(k-1)}
\end{equation}
which can also be written as a functional equation over $\F_+$ as
\begin{equation}\label{b57}
	H^\times |z^\gamma_R\rangle^{(k)}=z_R|z^\gamma_R\rangle^{(k)}
	+k|z^\gamma_R\rangle^{(k-1)};\hspace{.3in}k=0,1,\dots,r-1\;.
\end{equation}

If we use the normalization of the basis vectors defined
in~(\ref{bohm40}), and write~(\ref{b57}) out in detail then we obtain
\begin{align}
	H^\times |z^-_R\mer^{(0)} &= z_R\, |z^-_R\mer^{(0)} \nonumber \\
	H^\times |z^-_R\mer^{(1)} &= z_R\, |z^-_R\mer^{(1)} +\Gamma\,
	|z^-_R\mer^{(0)} \nonumber \\
	&~~\vdots  		\label{b58}\\
	H^\times |z^-_R\mer^{(k)} &= z_R\, |z^-_R\mer^{(k)} +\Gamma\,
	|z^-_R\mer^{(k-1)} \nonumber \\
	&~~\vdots  \nonumber \\
	H^\times |z^-_R\mer^{(r-1)} &= z_R\, |z^-_R\mer^{(r-1)} +\Gamma\,
	|z^-_R \mer^{(r-2)} \; .\nonumber 
\end{align}
(and the same for $|z^\gamma_R\mer^{(k)}$).
This means that $H^\times$ restricted to the subspace ${\cal M}_{z_R}$
is a Jordan operator of degree $r$ (in the standard notation the
operator $\frac{1}{\Gamma}H^\times$ is the Jordan operator of degree $r$), and
the vectors $|z^\gamma_R\mer^{(k)},~k=0,~1,~2,\dots,~r-1$ are Jordan
vectors of degree $k+1$.~\cite{baumgartel-gantmacher-lancaster}
They fulfill the generalized eigenvector equation~\cite{eigenvector}
\begin{equation}\nonumber
	\hspace{5.6cm}(H^\times -z_R)^{k+1}|z_R^-\mer^{(k)} =0\;.\hspace{4.8cm}
	(\ref{b58}')
\end{equation}

We write the equations~(\ref{b58}) again in the form~(\ref{b56}) and
arrange them as a matrix equation. Since the basis system includes,
according to~(\ref{bohm44}), in addition to the
$|z_R^\gamma\mer^{(k)}$, $k=0,1,2,\dots,r-1$, also the
$|E^-\rangle,~-\infty_{\rm II}<E\leq0$, we indicate this by a
continuously infinite diagonal matrix equation which we write as:
\begin{equation}\label{b59}
	\Bigl(\langle H\psi^-|E^-\rangle\Bigr)=
	\Bigl(\langle\psi^-|H|E^-\rangle\Bigr)=
	\Bigl(E\Bigr)\Bigl(\langle\psi^-|E^-\rangle\Bigr) 
\end{equation}
where $\left(\langle\psi^-|E^-\rangle\right)$ indicates a continuously
infinite column matrix. Then~(\ref{b58}) and~(\ref{b59}) together can
be written in analogy to~(\ref{eq:5.62}) as:
\begin{eqnarray}\nonumber
	\left( \begin{matrix} \<H\psi^-|z^-_R\mer^{(0)}\\
                \<H\psi^-|z^-_R\mer^{(1)} \\
                   \;\vdots\\
		   \;\vdots\\
                \<H\psi^-|z^-_R\mer^{(r-1)} \\
         	\<H\psi^-|E^- \rangle
	\end{matrix} \right) 					
      &=&\left( \begin{matrix} \<\psi^-|\, H^\times \, |z^-_R\mer^{(0)} \\
                \<\psi^-|\, H^\times \, |z^-_R\mer^{(1)} \\
                   \;\vdots\\
		   \;\vdots\\
                \<\psi^-|\, H^\times \, |z^-_R\mer^{(r-1)} \\
         \<\psi^-|\, H^\times \, |E^- \rangle
	\end{matrix} \right)\\
      &&\label{eq:54}\label{paul49}\label{b60}\\
	\nonumber
      &=&\begin{pmatrix} z_R &0&0&\ldots&0&0\\
                \Gamma&z_R &0&\ldots&0&\\
                0&\Gamma&z_R&\ldots&0&\vdots\\       
                \vdots&\vdots&\ddots &\ddots&\vdots&\\
                0&0&\ldots&\Gamma&z_R &0\\
          0&&\ldots&&0&\left( E\right) 
	\end{pmatrix}
 	\begin{pmatrix} \<\psi^-|z^-_R\mer^{(0)}\\
                 \<\psi^-|z^-_R\mer^{(1)}\\
                    \;\vdots\\
                    \;\vdots\\
                 \<\psi^- |z^-_R\mer^{(r-1)}\\
                 \<\psi^- |E^-\rangle 
	\end{pmatrix}\hspace{.5in}
\end{eqnarray}
In this matrix representation of $H^\times$, the upper left
$r\times r$ submatrix associated with the complex
eigenvalue $z_R$ is a (lower) Jordan block of degree $r$.
We have chosen the Jordan vectors with the normalization of
(\ref{bohm40}) in order to obtain the Jordan block in a form closest
to the standard form, but with $\Gamma$'s in place of $1$'s on the subdiagonal.

It is instructive also to write down the adjoint (i.e. transposed
complex conjugate) of the matrix equation~(\ref{b60}) because it will
clarify the notation and display the upper Jordan block. Taking the
transposed and complex conjugate of~(\ref{b60}) we obtain:
\begin{eqnarray}
	&&\Bigl(\,^{(0)}\klam^-z_R|H|\psi^-\rangle,	
			\dots,
	      \,^{(r-1)}\klam^-z_R|H|\psi^-\rangle,
	      \,\langle^-E|H|\psi^-\rangle\Bigr)=\\
	&&\Bigl(\,^{(0)}\klam^-z_R|\psi^-\rangle,	
			\dots,
	      \,^{(r-1)}\klam^-z_R|\psi^-\rangle,
	      \,\langle^-E|\psi^-\rangle\Bigr)
	\left(\begin{array}{cccccc}
		z^*_R&\Gamma&0&\cdots&0&0\\
		0&z^*_R&\Gamma&&0&\\
		0&0&z^*_R&\ddots&\vdots&\vdots\\
		\vdots&\vdots&&\ddots&\Gamma&\\
		0&0&0&\cdots&z^*_R&0\\
		0&&\cdots&&0&(E)
	\end{array}\right)\nonumber
\end{eqnarray}

With the derivation of~(\ref{bohm44}) and~(\ref{b60}) we have reduced
the problem of finding the
vectors (and their properties) associated with the higher order poles of
the S-matrix to the spectral theory of finite dimensional (non-normal)
complex matrices, which is well documented in the mathematical
literature~\cite{baumgartel-gantmacher-lancaster}. If in addition to
the $r$-th order pole at $z_R$ there are other $r_i$-th order poles at
$z_{R_i}$, then for each of these poles we have to add another Jordan
block of degree $r_i$ to the matrix in~(\ref{b60}) .

\noindent We could now refer for further results to the mathematics
literature of $r\times r$ complex matrices, but we can also obtain
these results easily from~(\ref{b55}) and~(\ref{b55'}).

Applying to the right-hand side of~(\ref{b55}) the Leibniz rule we obtain
\begin{equation}\label{b61}
	\langle\psi^-|\, f(H^\times )\, |z^\gamma_R\rangle^{(k)}=
	\sum^{k}_{\nu=0}\begin{pmatrix}k\\\nu\end{pmatrix}\Bigl[ f^{(\nu)}(z)
	\left(\langle\psi^- |z^\gamma\rangle\right)^{(k-\nu)}\Bigr]_{z=z_R}
\end{equation}
where $f^{(\nu)}(z)$ is the $\nu$-th derivative of the holomorphic
function $f(z)$ with respect to $z$ and 
$\langle\psi^-|z^\gamma\rangle^{(k-\nu)}\equiv
\left(\langle\psi^- |z^\gamma\rangle\right)^{(k-\nu)}$ is the $(k-\nu)$-th
derivative of $\langle\psi^- |z^\gamma\rangle$. We now
insert~(\ref{bohm40}) on both sides of~(\ref{b61}) and obtain:
\begin{equation}\label{b62}
	\frac{k!}{\Gamma^k}\langle\psi^-|\, f(H^\times )\,
	|z^\gamma_R\mer^{(k)}=\sum^{k}_{\nu=0}
	\frac{k!}{\nu!(k-\nu)!}f^{(\nu)}\!(z_R)\;\langle\psi^-
	|z^\gamma_R\mer^{(k-\nu)}\frac{(k-\nu)!}{\Gamma^{k-\nu}}
\end{equation}
From this we obtain
\begin{equation}\label{b63a}
	\langle\psi^-|\, f(H^\times )\,|z^\gamma_R\mer^{(k)}
	\;=\;\sum^{k}_{\nu=0}\frac{\Gamma^\nu}{\nu!}\;f^{(\nu)}(z_R)\;
	\langle\psi^-|z^\gamma_R\mer^{(k-\nu)}
\end{equation}
or as a functional equation:
\begin{equation}\label{b63b}
	f(H^\times )\,|z^\gamma_R\mer^{(k)}=\sum^{k}_{\nu=0}\frac{\Gamma^\nu}
	{\nu!}f^{(\nu)}(z_R)|z^\gamma_R\mer^{(k-\nu)}
\end{equation}
(Note in this calculation that $\langle\psi^-|z^-\mer^{(n)}$ 
is {\em not} the {$n$-th} derivative of
$\langle\psi^-|z^-\mer^{(0)}$, whereas
$\langle\psi^-|z^-\rangle^{(n)}$ is the $n$-th derivative of
$\langle\psi^-|z^-\rangle\in{{\cal S}\cap{\cal H}_+^2}$.
Therefore it is better to work with the $|z^-\rangle^{(k)}$ than 
with the $|z^-\mer^{(k)}$.)

In the theory of finite dimensional Jordan 
operators~\cite{baumgartel-gantmacher-lancaster}, the
equality~(\ref{b63b}) is often called the Lagrange-Sylvester formula 
and is written as a matrix equation (using lower Jordan blocks 
for $H^\times$ as in~(\ref{b60})):
\begin{align}\label{b64}
	&\begin{pmatrix}
		\langle\psi^-|\, f(H^\times )\, |z^-_R\mer^{(0)} \\
		\langle\psi^-|\, f(H^\times )\, |z^-_R\mer^{(1)} \\
		\quad\vdots \\
		\quad\vdots\\
		\langle\psi^-|\, f(H^\times )\, |z^-_R\mer^{(r-1)} \\
	\end{pmatrix} = \\
      &=\begin{pmatrix} 
   		f(z) &0&\ldots&\ldots&0 \\
     		\frac{\Gamma}{1!}f^{(1)}(z)&f(z)&0&\ldots&0\\
 		\frac{\Gamma^2}{2!}f^{(2)}(z)&
   		\frac{\Gamma}{1!}f^{(1)}(z)&\ddots&\ddots&\vdots \\
   		\vdots&\vdots&\ddots&f(z)&0 \\
 		\frac{\Gamma^{r-1}}{(r-1)!}f^{(r-1)}(z)&
    		\frac{\Gamma^{r-2}}{(r-2)!}f^{(r-2)}(z)&
		\ldots&\frac{\Gamma}{1!}f^{(1)}(z)&f(z)
	\end{pmatrix}_{\!\!z=z_R}
	\begin{pmatrix}
		\langle\psi^-|z^-_R \mer^{(0)} \\
		\langle\psi^-|z^-_R \mer^{(1)} \\
  		\quad \vdots \\
  		\quad \vdots \\
		\langle\psi^-|z^-_R \mer^{(r-1)} \\
	\end{pmatrix}\nonumber
\end{align}
The $r\times r$ submatrix equation of~(\ref{b60}) is a special case of
this for $f(H^\times)=H^\times$. Equation~(\ref{b64}) is not the
complete matrix representation of $f(H^\times)$, because the infinite
diagonal submatrix due to the first term in~(\ref{bohm44}),
\begin{equation}\label{b65}
	\Bigl(\langle f^*(H)\psi^-|E^+\rangle\Bigr)=\Bigl(E\Bigr)\Bigl(
	\langle\psi^-|E^+\rangle\Bigr),\hspace{.3in}-\infty_{\rm
	II}<E\leq 0\;, 
\end{equation}
has been omitted. Equation~(\ref{b64}) gives the restriction of
$f(H^\times)$ to the $r$-dimensional subspace ${\cal M}_{z_R}\subset\Phi^\times_+$.

The function of $H^\times$ that we are particularly interested in is
the time evolution operator $f(H^\times)=e^{-iH^\times t}_+$. 
It can be defined in $\F^\times_+$ only for those values of the 
parameter $t$ for which $e^{iHt}:\F_+ \longrightarrow \F_+$ is a 
$\tau_{\F_+}$-continuous operator.  This is the case for $t\ge 0$, but
not for $t\le 0$.  (For $\langle \psi^-|E^-\rangle\in\CS\cap\HH^2_-$, the
function $\langle e^{iHt}\psi^- |E^-\rangle = e^{-iEt}\langle\psi^-
|E^-\rangle$ is an element of $\CS\cap\HH^2_-$ only for $t\ge 0$.)
Thus, for $t\ge 0$, we can use (\ref{b63b}) with $f(z)=e^{-izt}$ and
$f^{(\nu)}(z)=(-it)^\nu e^{-izt}$, and we obtain the following functional equation in $\CM_{z_R} \subset\F^\times_+$:
\begin{equation}\label{b66}
	e^{-iH^\times t}|z^\gamma_R\mer^{(k)}\,=\,e^{-iz_Rt}\sum_{\nu=0}^k
	\frac{\Gamma^\nu}{\nu!}(-it)^\nu|z^\gamma_R\mer^{(k-\nu)}\;.
\end{equation}
In terms of the vectors $|z^\gamma_R\rangle^{(k)}$ this can be 
written (using~(\ref{bohm40})):
\begin{subeqnarray}\label{b67}\label{eq:59}\label{timeevo}
	e^{-iH^\times t}|z^\gamma_R\rangle^{(k)}=e^{-iz_Rt}
	\sum^k_{\nu=0}\begin{pmatrix}k\\\nu\end{pmatrix}\,
	(-it)^\nu\,|z^\gamma\rangle^{(k-\nu)} 
\end{subeqnarray}
or taking the complex conjugate (in analogy to going from~(\ref{bohm30}a) 
to~(\ref{bohm30}b)):
\begin{equation}\nonumber
	\hspace{3.5cm}\,^{(k)}\langle^\gamma z_R|e^{iHt}=
	e^{iz^*_Rt}\sum^k_{\nu=0}\begin{pmatrix}k\\\nu\end{pmatrix}\,(it)^\nu
	\;\,^{(k-\nu)}\langle^\gamma
	z_R|\;.\hspace{3.4cm}(\ref{b67}{\rm b})
\end{equation}
The vectors $|z^\gamma_R\rangle^{(k)}$ in the above equations can be replaced by the vectors $|z^-_R\rangle^{(k)}$.

It is important to note that the time evolution $e^{-iH^\times t}$
transforms between different $|z^-_R\mer^{(k)}\, ,\; k=1 ,\, 2
\ldots, n$, that belong to the same pole of order $r$ at $z=z_R$,
but the time evolution does not transform out of $\CM_{z_R}$.
On the basis vectors $|E^+\rangle$ of the first term in~(\ref{bohm44}) 
the time evolution is diagonal 
\begin{equation}\label{b68}
	e^{-iH^\times t}|E^+\rangle=e^{-iEt}|E^+\rangle\;.
\end{equation}
The equation~(\ref{b66}) and~(\ref{b67}) can be written as a matrix 
equation on the subspace {${\CM_{z_R}\subset\F_+}$}:
\begin{align}\label{b69}&
	\begin{pmatrix}
		\langle\psi^-|\, e^{-iH^\times t}\, |z^-_R\mer^{(0)} \\
		\langle\psi^-|\, e^{-iH^\times t}\, |z^-_R\mer^{(1)} \\
		\quad\vdots \\
		\quad\vdots\\
		\langle\psi^-|\, e^{-iH^\times t}\, |z^-_R\mer^{(r-1)}
	\end{pmatrix} =
\\&=
        \begin{pmatrix} e^{-iz_Rt} &0&\ldots&\ldots&0 
	\\	\frac{(-it\Gamma)}{1!}e^{-iz_Rt}&e^{-iz_Rt}&0&\ldots&0
	\\ 	\frac{(-it\Gamma)^2}{2!}e^{-iz_Rt}&
		\frac{(-it\Gamma)}{1!}e^{-iz_Rt}&\ddots&\ddots&\vdots 
	\\	\vdots &\vdots&\ddots&e^{-iz_R t}&0 
	\\	\frac{(-it\Gamma)^{r-1}}{(r-1)!}e^{-iz_Rt}&
        	\frac{(-it\Gamma)^{r-2}}{(r-2)!}e^{-iz_Rt}&\ldots& 
         	\frac{(-it\Gamma)}{1!}e^{-iz_R t}&e^{-iz_R t}  
  	\end{pmatrix} 
	\begin{pmatrix}
		\langle\psi^-|z^-_R \mer^{(0)} \\
		\langle\psi^-|z^-_R \mer^{(1)} \\
  		\quad \vdots\\
  		\quad \vdots\\
		\langle\psi^-|z^-_R \mer^{(r-1)}
  	\end{pmatrix}\nonumber   
\end{align}

As an example let us consider the special case of a double pole, $r=2$,
$k=0,~1$. The formula~(\ref{b66}) for the zeroth order Gamow vector is then
\begin{equation}\label{b70a}
	e^{-iH^\times t}|z^-_R\mer^{(0)}= e^{-iE_Rt}e^{-(\Gamma_R/2)t}
	|z^-_R\mer^{(0)}, \hspace{.3in} t\geq0,  
\end{equation}
and for the first order Gamow vector it is
\begin{equation}\label{b70b}
	e^{-iH^\times t}|z^-_R\mer^{(1)} = e^{-iz_Rt}\left( 
	|z^-_R\mer^{(1)}+(-it\Gamma)|z^-_R\mer^{(0)}\right) .
\end{equation}

It has been known for a long time~\cite{polynomial} that a double pole and
in general all higher order S-matrix poles lead to a polynomial time
dependence in addition to the exponential. However, it was not clear
what the vectors were that have such a time evolution. Here we have
seen that they are Jordan vectors of degree $r$ or less,
and that they are Gamow vectors, $|z^\gamma_R\mer^{(k)}\in\Phi^\times_+$. 
We have also shown that the time evolution operator is not diagonal in the
basis~(\ref{bohm38}) but transforms a Gamow-Jordan vector of degree
$(k+1)$ into a superposition~(\ref{b66}) of Gamow-Jordan vectors of 
the same and all lower degrees with a time dependence $t^\nu$ in 
addition to the exponential that depends upon the degrees of the 
resulting Gamow-Jordan vectors. 

\section{Possible Physical Interpretations of the Gamow-Jordan Vectors}
\label{sec:possible}
\setcounter{equation}{0}

In the previous section we defined the higher order Gamow vectors
$|z^-_R\rangle^{(k)}\in\Phi^\times_+$ from the $r$-th order pole term
of a unitary S-matrix. We showed that they are the discrete members of
a complete basis system for the vectors
$\phi^+\in\Phi_-$,~(\ref{bohm44}), and we derived their mathematical
properties: In~(\ref{b57}) and~(\ref{b58}), we showed that they are 
Jordan vectors of degree $k+1$, and in~(\ref{b64})
and~(\ref{b69}), we obtained the Lagrange-Sylvester formula and the
time evolution.

The mathematical procedure that we used for the $r$-th order pole term
is a straightforward generalization of the definitions and derivations
that had been used for an ordinary, zeroth order Gamow vector and
first order poles of the S-matrix~\cite{bohm-group}.

Gamow states of zeroth order with their empirically well established
properties (exponential time evolution, Breit-Wigner energy
distribution) have been abundantly observed in nature as resonances
and decaying states. Theoretically, there is no reason why the other
quasistationary states (i.e. states that also cause large time delay
in a scattering process (\cite{bohm} sect.~XVIII.6) and are associated
with integers $r>1$ in~(\ref{b16a})) should not exist. However, no such
quasistationary states have so far been established empirically. One
argument against their existence was that the polynomial time dependence,
that was always vaguely associated with higher order
poles~\cite{polynomial}, has not been observed for quasistationary states.

The question that we want to discuss in this section is, whether there is
an analogous physical interpretation for the higher order Gamow vectors
as for the ordinary Gamow vectors, namely as states
which decay (for $t>0$) or grow (for $t<0$) in one prefered direction
of time (``arrow of time'') and obey the exponential law.
Since we have now well defined vectors associated with an $r$-th order
pole, we can attempt to define physical states which have well defined
properties that can be tested experimentally. 

In this section we are
dealing with physical questions about hypothetical objects associated
with the $r$-th order pole. We therefore have first to conjecture the higher
order Gamow state, before we can derive their properties. We start
with the known cases.

In von Neumann's definition of a pure stationary state one uses a
dyadic product $W=|f\rangle\langle f|$ of energy eigenvectors
$|f\rangle$ in Hilbert space. In analogy to this, microphysical
Gamow states connected with first order poles have been defined 
as dyadic products of zeroth order Gamow
vectors~\cite{bohm}~\cite{bohm-gadella-maxson}:
\begin{equation}\label{g60}
	W^G=|\psi^G\rangle\langle\psi^G| 
	=|z^-_R\rangle\langle^-z_R|\, \equiv \, W^{(0)}
\end{equation}
(Since for the generalized vectors $|\psi^G\rangle=\sqrt{2\pi\Gamma}
\, |z^-_R\rangle$ or $|z^-_R\rangle$ we cannot talk of normalization
in the ordinary sense, it is not important at this stage whether or
not to use the ``normalization'' factor of $2\pi\Gamma$ 
in $W^G$.

\noindent The time evolution of the Gamow state~(\ref{g60}) is then
given according to~(\ref{bohm30}) by:  
\begin{eqnarray}
	W^G(t)&\equiv&e^{-iH^\times t}\,|\psi^G\rangle
	\langle\psi^G|\,e^{iHt} \nonumber\\ 		\label{WG62}\label{g61}
	&=&e^{-iz_R
	t}\,|\psi^G\rangle\langle\psi^G|\,e^{iz^*_Rt}\\
 	&=&e^{-i(E_R-i(\Gamma/2))t}|\psi^G\rangle\langle\psi^G| 
	\,e^{i(E_R+i(\Gamma/2))t}\nonumber\\
	&=&e^{-\Gamma t}\,W^G(0)\;,\hspace{4.5cm}t\geq0\;.\nonumber 
\end{eqnarray}
Mathematically, the equation~(\ref{g61}) is to be understood as a
functional equation like~(\ref{bohm28}) and~(\ref{bohm29}):
\begin{subeqnarray}\label{g62}\label{WG63}
	\langle\psi^-_1|W^G(t)|\psi^-_2\rangle
	&=&e^{-\Gamma
	t}\langle\psi^-_1|W^G|\psi^-_2\rangle\hspace{1cm}{\rm or}\\
	\langle\psi^-|W^G(t)|\psi^-\rangle
	&=&e^{-\Gamma t}\langle\psi^-|W^G|\psi^-\rangle\\
	&&{\rm for~all}\quad\psi^-,\psi^-_1,\psi^-_2\in\Phi_+
	\quad{\rm and}\quad t\geq0.\nonumber
\end{subeqnarray}

The mathematical form~(\ref{g62}) of the time evolution of $W^G$
shows how important it is in our RHS formulation to know what
question one wants to ask about a Gamow state when one makes the
hypothesis~(\ref{g60}). The vectors $\psi^-\in\Phi^+$ represent
observables defined by the detector (registration apparatus). The
operator $W^G$ represents the microsystem that affects the
detector. Therefore the quantity $\langle\psi^-|W^G|\psi^-\rangle$ is
the answer to the question: What is the probability that the
microsystem affects the detector?


If the detector is triggered at a later time
$t$, i.e. when the observable has been time translated 
\begin{equation}\label{g63}
	|\psi^-\rangle\langle\psi^-|\quad\longrightarrow\quad
	e^{iHt}|\psi^-\rangle\langle\psi^-|e^{-iHt}\,
	=\,|\psi^-(t)\rangle\langle\psi^-(t)|
\end{equation}
then the same question for $t\geq0$ has the answer: The probability
that the microsystem affects the detector at $t>0$ is
\begin{eqnarray}\nonumber
	\hspace{-.3in}\langle\psi^-(t)|W^G|\psi^-(t)\rangle
	&=&\langle e^{-iHt}\psi^-|W^G|e^{iHt}\psi^-\rangle\\\label{g64}
	&=&\langle\psi^-|e^{-iH^\times t}W^Ge^{iHt}|\psi^-\rangle\\\nonumber
	&=&e^{-\Gamma t} \langle \psi^-| W^G |\psi^-\rangle\;.
\end{eqnarray}
This means that~(\ref{g64}) is the probability to observe the decaying
microstate at time $t$ relative to the probability
$\langle\psi^-|W^G|\psi^-\rangle$ at $t=0$, (which one can
``normalize'' to unity by choosing an appropriate factor on the
right-hand side of~(\ref{g60})).

The question that one asks in the scattering experiment of fig.~1 is
different. There the pole term (P.T.) of~(\ref{eq:regular}) 
describes how the microsystem propagates the effect which the
preparation apparatus (accelerator, described by the state $\phi^+$)
causes on the registration apparatus (detector, described by the
observable $\psi^-$).

In conventional orthodox quantum theory one only deals with ensembles
and with observables measured on ensembles. Their mathematical representations,
e.g., $|\phi\rangle\langle\phi|$ for the state of the ensemble and 
$\, |\psi\rangle\langle\psi|\,$ for the observable, are from the same
space $\bf{\Phi}$, i.e. $\phi,~\psi\in\Phi$. (And if one is
mathematically precise then one chooses for $\Phi$ the Hilbert
space, $\Phi={\cal H}$.) On this level, one cannot talk
of single microsystems, and there are no mathematical objects in
orthodox quantum mechanics to describe a single microsystem. 
Still, it is intuitively attractive to imagine that the
effect by which the preparation apparatus acts on the registration
apparatus is carried by single physical entities, the
microphysical systems~\cite{ludwig-foundations}. 

According to the physical interpretation of the RHS
formulation, ``real'' physical entities connected with an experimental
apparatus, like the states $\phi$ defined by the preparation apparatus
or the property $\psi$ defined by the registration apparatus, are
assumed to be elements of $\Phi$, but states and observables are
distinct. In particular, states and observables of a scattering 
experiment are distinct and described by $\Phi_-$ of~(\ref{bohm6}a)
and $\Phi_+$ of~(\ref{bohm6}b). However,
mathematical entities describing microphysical systems are not assumed
to be in $\Phi$. The energy distribution for a microphysical system does not
have to be a well-behaved (continuous, smooth, rapidly decreasing)
function of the physical values of the energy $E$, like the functions
$\langle E|\psi\rangle$ describing the energy resolution of the
detector, or the functions $\langle E|\phi\rangle$ describing the energy
distribution of the beam. Hence, for the hypothetical entities
connected with microphysical systems, like Dirac's ``scattering
states'' $|{\rm {\bf p}}\rangle$ or Gamow's ``decaying states''
$|E-i\Gamma/2\rangle$, the RHS formulation uses elements of
$\Phi^\times$, $\Phi^\times_+$, and 
$\Phi^\times_-$~\cite{cologne,bohm-gadella-maxson}. 
The time evolution  of the ``state'' vectors
for the decaying microphysical systems, e.g.~(\ref{eq:5.89prime})
or~(\ref{g61}), can 
be obtained from the well established time evolution of the quantum
mechanical observable~(\ref{g63}) using the definition of the conjugate
operator as in~(\ref{bohm28}).

Because of the difference between {$\psi^-\in\Phi_+$} for the
observables and {$\phi^+\in\Phi_-$} for the prepared states one needs a different mathematical
description for the same microphysical state, depending upon the
question one is asking. If one asks the question with what
probability the microphysical state affects the detector
$\psi^-(t)$, then the microphysical
state is described by~(\ref{g60}). In a resonance scattering
experiment of fig.~1 one asks another question: What is the
probability to observe $\psi^-(t)$ in a microphysical resonance state
of a scattering experiment with the prepared in-state $\phi^+$?

\noindent In distinction to a decay experiment, where one just asks
for the probability of $\psi^-\in\Phi_+$, in the resonance scattering
experiment one asks for the probability that relates
$\psi^-\in\Phi_+$ to $\phi^+\in\Phi_-$ via the microphysical resonance
state. Therefore the mathematical quantity that describes the
microphysical resonance state in a scattering experiment cannot be
given by $|z^-_R\rangle\langle^-z_R|$ of~(\ref{g60}), but must be
given by something like $|z^-_R\rangle\langle^+z_R|$.

\noindent The probability to observe $\psi^-$ in the prepared state
$\phi^+$, independently of how the effect of $\phi^+$ is carried to
the detector $\psi^-$ is given by the S-matrix element~(\ref{bohm17}),
$|(\psi^-,\phi^+)|^2$. The probability amplitude that this effect is
carried by the microphysical resonance state is then given by the pole
term $(\psi^-,\phi^+)_{\rm P.T.}$, equation~(\ref{bohmprime}).

In analogy to~(\ref{g64}) one can now also compare these
probabilities at different times. For this purpose 
one translates the observable $\psi^-$ in the pole
term~(\ref{eq:regular}) in time by an amount $t\geq 0$,
\begin{equation}\label{g65}
	\psi^-\longrightarrow\psi^-(t)=e^{iHt}\psi^-;\hspace{.3in}t\geq0
\end{equation}
(which corresponds to turning on the detector at a time $t\geq0$ later
than for $\psi^-$ ). One obtains
\begin{align}\nonumber
	\left(\psi^-(t),\phi^+\right)_{\rm P.T.} &
	=\,-2\pi\Gamma\,\langle e^{iHt}\psi^-|z_R^-\rangle 
	\langle^+z_R|\phi^+\rangle\\ \label{WG66a}&
	=\,-2\pi\Gamma \, \langle \psi^-|e^{-iH^\times t}
	|z_R^-\rangle \langle^+z_R|\phi^+\rangle\\&	
	=\,-2\pi\Gamma \, e^{-iz_Rt}\langle \psi^-| z_R^-\rangle 
	\langle^+ z_R | \phi^+ \rangle\nonumber\\&\nonumber
	= e^{-iE_Rt}e^{-\Gamma t/2}\left(\psi^-,\phi^+\right)_{\rm P.T.}
\end{align}
This means that the time dependent probability, due to the first order
pole term, to measure the observable $\psi^-(t)$ in the state $\phi^+$
is given by the exponential law:
\begin{equation}\label{WG66b}
	|(e^{iHt}\psi^-,\phi^+)_{\rm P.T.}|^2 
	=e^{-\Gamma t} \, |(\psi^-,\phi^+)_{\rm P.T.}|^2. 
\end{equation}
This is as one would expect it if the action of the preparation
apparatus on the registration apparatus is carried by an
exponentially decaying microsystem (resonance) described by
a Gamow vector.

Therewith we have seen that there are two ways  in which a resonance
can appear in experiments and therefore there should be two forms of
representing the decaying Gamow state (for the case $r=1$ so
far)\footnote{An analogous statement holds for the Gamow states 
associated with the pole in the upper half-plane.}
\begin{subeqnarray}\label{68a}\label{g68a}
	{\rm by}\hspace{.8cm}&|z^-_R\rangle\langle^+z_R|
	&{\rm~in~a~scattering~experiment,}\\\label{68b}\label{g68b}
	{\rm and~by}&|z^-_R\rangle\langle^-z_R|&
	{\rm~in~a~decay~experiment.}
\end{subeqnarray}
The first representation is the one used in the S-matrix when one
calculates the cross section; the second representation is the 
one used when one calculates the Golden Rule (decay rate). 
In contrast to von Neumann's formulation where a given state
(representing an ensemble prepared by the preparation apparatus) is
always described by one and the same density operator $W$, the
representation of the microphysical state in the RHS formulation depends upon
the question one asks, i.e. upon the kind of experiment which
one wants to perform.
That a theory of the microsystems must include the methods of the
experiments has previously been emphasized in~\cite{ludwig-foundations}.

After this preparation we are now ready to conjecture the mathematical
representation of a higher order Gamow state (a quasistationary state
with $r>1$). 

In analogy to the correspondence between~(\ref{68a}a) and~(\ref{68b}b) we
conjecture that for the case of general $r$ we have also two distinct
representations of the Gamow state. The one for resonance scattering 
is already determined as in the case for $r=1$ by the (negative of the)
pole 
term~(\ref{bohm41}), and is therefore given~by 
\begin{subeqnarray}\label{g69a}
	W_{\rm P.T.}&=&-2\pi\Gamma\sum_{n=0}^{r-1}\begin{pmatrix}r\\
	n+1\end{pmatrix}(-i)^n\frac{\Gamma^n}{n!}\sum_{k=0}^n
	\begin{pmatrix}n\\k\end{pmatrix}|z^-_R\rangle^{(k)}
	\;^{(n-k)}\langle^+z_R|\\
	&=&-2\pi\Gamma\sum_{n=0}^{r-1}\begin{pmatrix}r\\n+1\end{pmatrix}
	(-i)^nW^{(n)}_{\rm P.T.}\nonumber
\end{subeqnarray} 
where we have used the operator defined in~(\ref{bohm42}):
\begin{subeqnarray}\label{g69b}
	W^{(n)}_{\rm P.T.}=\frac{\Gamma^n}{n!}\sum_{k=0}^n\begin{pmatrix}n\\
	k\end{pmatrix}|z^-_R\rangle^{(k)}\;^{(n-k)}\langle^+z_R|
	=\sum_{k=0}^n|z^-_R\mer^{(k)}\;^{(n-k)}\klam^+\!\!z_R|\;.
\end{subeqnarray}

In analogy to~(\ref{g68b}b) we would then conjecture that the $r$-th
order microphysical decaying state is described by the state operator
\begin{eqnarray}
	\hspace{2.8cm}W&=&2\pi\Gamma\sum_{n=0}^{r-1}\begin{pmatrix}
	r\\n+1\end{pmatrix}(-i)^n\frac{\Gamma^n}{n!}\sum_{k=0}^n
	\begin{pmatrix}n\\k\end{pmatrix}|z^-_R\rangle^{(k)}\;^{(n-k)}
	\langle^-z_R|\hspace{1.7cm}(\ref{g69a}{\rm b})\nonumber\\
	&=&2\pi\Gamma\sum_{n=0}^{r-1}\begin{pmatrix}r\\n+1\end{pmatrix}
	(-i)^nW^{(n)}\nonumber 
\end{eqnarray}
(up to a normalization factor which will have to be determined by
normalizing the overall probability to 1).
Since~(\ref{g68b}b) is postulated to be the zeroth order Gamow
state representing a resonance, (\ref{g69a}b) 
is conjectured to be the $r$-th order Gamow state.\footnote{
We want to mention that mathematically there is an important
difference between~(\ref{g69a}a) and~(\ref{g69a}b) because
$\langle\psi^-|z^-\rangle^{(k)}\,^{(n-k)}\langle^+z|\phi^+\rangle$ 
are analytic functions for $z$ in the lower half-plane, whereas the 
$\langle\psi^-_1|z^-\rangle^{(k)}\,^{(n-k)}\langle^-z|\psi^-_2\rangle$ are
not.}

Whether the microphysical state of the (hypothetical) quasistationary
microphysical system is always represented by the mathematical
object~(\ref{g69a}b) or whether also each individual 
\begin{eqnarray}\nonumber
	\hspace{2.3cm}W^{(n)}&=&\frac{\Gamma^n}{n!}\sum_{k=0}^n
	\begin{pmatrix}n\\k\end{pmatrix}|z^-_R\rangle^{(k)}\;^{(n-k)}
	\langle^-z_R|=\sum_{k=0}^n|z^-_R\mer^{(k)}\;^{(n-k)}\klam^-z_R|\;;\\
	&&\hspace{3.45cm}n=0,1,\dots,r-1\;,\hspace{4cm}
	(\ref{g69b}{\rm b})\nonumber
\end{eqnarray}
has a separate physical meaning, is not clear. So far it is not even
certain that higher order poles of the S-matrix describe
anything in nature (though there are no theoretical reasons that
exclude these isolated singularities of the S-matrix.)
But if these hypothetical objects do exist, the $r$-th order 
pole is associated with a mixed 
state~(\ref{g69a}b) whose irreducible components are given
by~(\ref{g69b}b). E.g., for the case $r=2$ (second order pole at $z_R$) we have:
\begin{equation}\label{g72}
	W^{(0)}=|z^-_R\rangle^{(0)}\;^{(0)}\langle^-z_R|\hspace{8cm}
\end{equation}
and
\begin{equation}\label{g73}
	W^{(1)}=\Gamma\Bigl( |z^-_R\rangle^{(0)}\;^{(1)}\langle^-z_R|+
	|z^-_R\rangle^{(1)}\;^{(0)}\langle^-z_R|\Bigr)\hspace{4.2cm}
\end{equation}
and
\begin{equation}\label{g74}
	W=2\pi\Gamma\Bigl(|z^-_R\rangle^{(0)}\;^{(0)}\langle^-z_R|-
	2i\Gamma\bigl(|z^-_R\rangle^{(0)}\;^{(1)}\langle^-z_R|+
	|z^-_R\rangle^{(1)}\;^{(0)}\langle^-z_R|\bigr)\Bigr)
\end{equation}
This means that the conjectural physical state associated with the
$r$-th order pole is a mixed state $W$, all of whose components
$W^{(n)}$, except for the zeroth component $W^{(0)}$, cannot be
reduced further into ``pure'' states given by dyadic products like
$|z^-_R\rangle^{(k)}\,^{(k)}\langle^-z_R|$. This is quite consistent
with our earlier remark that the label $k$ is not a quantum number
connected with an observable (like the suppressed labels
$b_2,\dots,b_n$). Therefore a ``pure state'' with a definite value of
$k$, like $|z^-_R\rangle^{(k)}\,^{(k)}\langle^-z_R|$, $k\geq1$, does
not make sense physically. A physical interpretation could only be
given to the whole $r$-dimensional space ${\cal M}_{z_R}$, 
(\ref{bohm48}). The individual $W^{(n)}$, {$n=0,~1,~2,\dots,r-1$}, act in the
subspaces ${\cal M}^{(n)}_{z_R}\subset{\cal M}_{z_R}$ which are spanned
by Gamow vectors of order $0,~1,\dots,n\,$ (Jordan vectors of degree
$n+1$, i.e. $(H^\times-z_R)^{n+1}{\cal M}^{(n)}_{z_R}=0$). There the
question is, whether
there could be a physical meaning to each $W^{(n)}$ separately, or
whether only the particular mixture $W$ given by~(\ref{g69a}b) can occur
physically.

Though the quantities $|z^-_R\rangle^{(k)}\,^{(k)}\langle^-z_R|$ will
have no physical meaning, even if higher order poles exist, they have
been considered~\cite{antoniou-gadella} and their time evolution is
calculated in a straightforward way from~(\ref{b66}):
\begin{eqnarray}\label{purestate2}\label{g75}
	&&\hspace{-1cm}e^{-iH^\times t}|z^-_R\mer^{(k)}\,^{(k)}
	\klam^-z_R|e^{iHt}=\\
	&&=e^{-\Gamma t}\sum^k_{l=0}\,\sum^k_{m=0}
	\frac{1}{l!} 
	\frac{1}{m!}
	(-it\Gamma)^l(it\Gamma)^m
	|z^-_R\mer^{(k-l)}\,^{(k-m)}\klam^-z_R |\;.\nonumber
\end{eqnarray}
This time dependence (as well as the time dependence
in~(\ref{b66})) is reminiscent of eq.~(\ref{bohm45}) in the 
reference of M.~L. Goldberger and K.~M. Watson~\cite{polynomial}. 

It shows the additional polynomial time dependence, that has always
been considered an obstacle to the use of higher order poles for
quasistationary states. A polynomial time dependence of this
magnitude (of the order of $\tau=\frac{1}{\Gamma}$) should have shown
up in many experiments.

We now derive the time evolution of the microphysical state operator
defined in~(\ref{g69b}b) using the time evolution obtained for
the Gamow-Jordan vector in~(\ref{b67}). It will turn out that this
operator, whose form was conjectured in analogy to the pole
term~(\ref{g69b}a), will have a purely exponential time
evolution. This was quite unexpected.

Inserting~(\ref{b67}a) and~(\ref{b67}b) into
\begin{eqnarray}\label{b75}
	W^{(n)}(t)=e^{-iH^\times t}W^{(n)}e^{iHt}
	= \frac{\Gamma^n}{n!}\sum_{k=0}^n \, 
	      	\begin{pmatrix} 
			n \\ k 
		\end{pmatrix} \,
	e^{-iH^\times t} |z^-_R \rangle^{(k)} \,^{(n-k)} 
	\langle^- z_R |\,e^{iHt}
\end{eqnarray}
we calculate:
\begin{eqnarray}
	W^{(n)}(t)&\hspace{-.5cm}&=e^{-iz_Rt}e^{iz^*_Rt}
	\frac{\Gamma^n}{n!}\sum_{k=0}^n
	\sum_{l=0}^k\sum_{m=0}^{n-k} 
	\begin{pmatrix} n \\ k \end{pmatrix}
	\begin{pmatrix} k \\ l \end{pmatrix}
	\begin{pmatrix} n\! -\! k \\ m \end{pmatrix}
	(-it)^{k-l}(it)^{n-k-m}
	|z^-_R\rangle^{(l)}\,^{(m)}\langle^- z_R|	\nonumber\\
	=&\hspace{-.5cm}&e^{-\Gamma t}\frac{\Gamma^n}{n!}\sum_{m=0}^n
	\sum_{l=0}^{n-m}\sum_{k=l}^{n-m} 
	\begin{pmatrix} n \\ k \end{pmatrix}
	\begin{pmatrix} k \\ l \end{pmatrix}
	\begin{pmatrix} n\! -\! k \\ m \end{pmatrix}
	(-it)^{k-l}(it)^{n-k-m} 
	|z^-_R\rangle^{(l)}\,^{(m)}\langle^- z_R|	\label{marc5.20}\\
	=&\hspace{-.5cm}&e^{-\Gamma t}\frac{\Gamma^n}{n!}\sum_{m=0}^n
	\sum_{l=0}^{n-m}\sum_{k=l}^{n-m} 
	\begin{pmatrix} n \\ m \end{pmatrix}
	\begin{pmatrix} \!n\!-m\!\\ l \end{pmatrix}
	\begin{pmatrix} \!n\!-\!m\!-\!l\\ k\!-\!l \end{pmatrix}
	(-it)^{k-l}(it)^{n-k-m} 
	|z^-_R\rangle^{(l)}\,^{(m)}\langle^- z_R|	\nonumber\\
	=&\hspace{-.5cm}&e^{-\Gamma t}\frac{\Gamma^n}{n!}\sum_{m=0}^n
	\begin{pmatrix} n \\ m \end{pmatrix}
	\sum_{l=0}^{n-m}\begin{pmatrix} \!n\!-m\!\\ l \end{pmatrix}
	|z^-_R\rangle^{(l)}\,^{(m)}\langle^- z_R|\sum_{k=l}^{n-m}
	\begin{pmatrix} \!n\!-\!m\!-\!l\\ k\!-\!l \end{pmatrix}
	(-it)^{k-l}(it)^{n-k-m}				\nonumber
\end{eqnarray}
In going from the second to the third line, the order of
summation has been changed, by keeping the same terms, as displayed in
fig.~3 for the case $n=3$. In going from the third to the fourth line
one uses the identity 
\begin{equation}\label{b79}
	\begin{pmatrix} n \\ k \end{pmatrix}
	\begin{pmatrix} k \\ l \end{pmatrix}
	\begin{pmatrix} n\! -\! k \\ m \end{pmatrix}=
	\begin{pmatrix} n \\ m \end{pmatrix}
	\begin{pmatrix} n\!-\!m\! \\ l \end{pmatrix}
	\begin{pmatrix} n\!-\!m\!-\!l \\ k\!-\!l \end{pmatrix}
\end{equation}
where
{\footnotesize{$\begin{pmatrix}n\\k\end{pmatrix}$}}$\equiv\frac{n!}{k!(n-k)!}$
are binomial coefficients.
\begin{figure}  
\begin{center}
\includegraphics[scale=0.7]{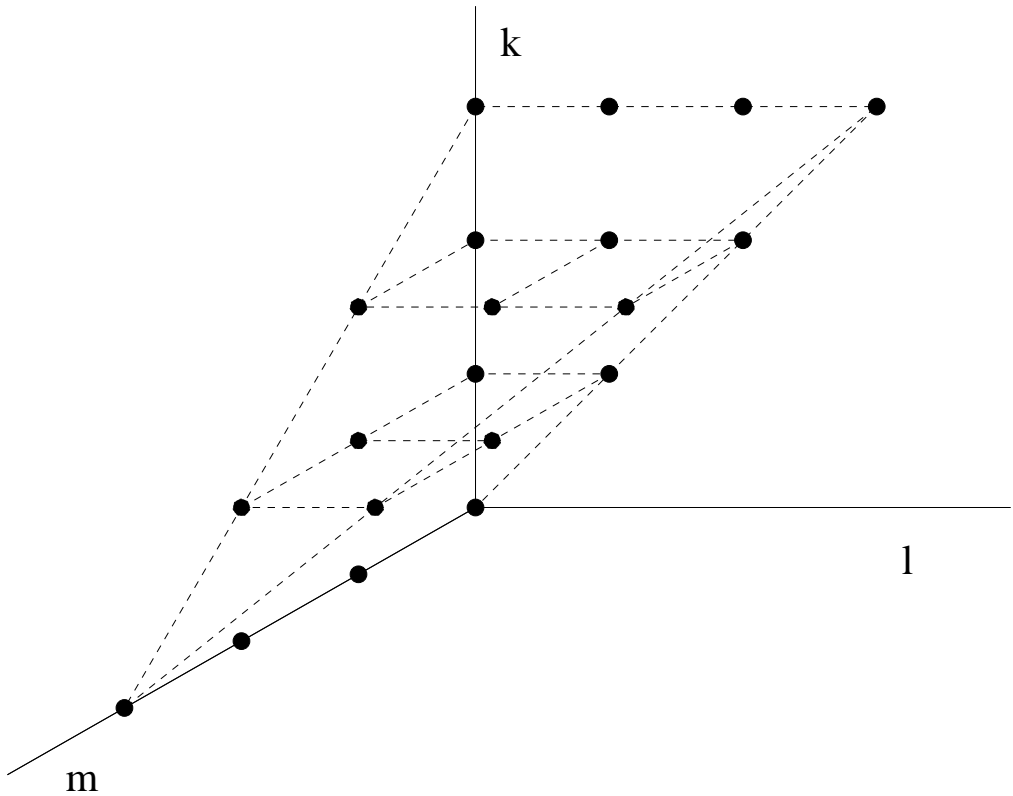}
\end{center}
        \caption{For the case $n=3$, the summation terms labeled by
        the parameters $k$, $l$, and $m$ are displayed as dots in the
        diagram to show that the summations of lines 2 and 3
        of~(\ref{marc5.20}) both contain the same terms.}
\end{figure}
Since the indices labeling the Gamow-Jordan vectors do not depend
upon $k$, the sum over $k$ may be performed using the binomial formula:
\begin{equation}\label{b82}
	\sum_{k=l}^{n-m}\, 
	\begin{pmatrix}\!n\!-\!m\!-\!l\!\\k\!-\!l\end{pmatrix}	
	(-it)^{k-l}\, (it)^{n-k-m}
	= (it-it)^{n-m-l} = 
	\left\{
		\begin{array}{c} 
			1 \quad \mbox{for $l=n-m$} \\ 
			0 \quad \mbox{for $l\neq n-m$} 
		\end{array} 
	\right\}=\delta_{l,n-m}
\end{equation}
Inserting~(\ref{b82}) into the fourth line of~(\ref{marc5.20}) and
performing the sum over $l$ then gives:
\begin{equation}\label{b82'}
	W^{(n)}(t)=e^{-\Gamma t}\,\frac{\Gamma^n}{n!}\,\sum_{m=0}^n 
      	\begin{pmatrix}n\\m\end{pmatrix}  
  	|z^-_R \rangle^{(n-m)} \,^{(m)} \langle^- z_R |
	=e^{-\Gamma t} \, W^{(n)}(0)\;; \hspace{.2in} t\geq 0 
\end{equation}

This means that the complicated non-reducible (i.e. ``mixed'')
microphysical state operator $W^{(n)}$ defined by~(\ref{g69b}b) 
has a simple purely exponential semigroup time
evolution, like the zeroth order Gamow 
state~(\ref{g68b}b). This operator is probably the
only operator formed by the dyadic products $|z^-_R \rangle^{(m)}
\,^{(l)} \langle^-z_R |$ with $~m,~l=0,~1,\cdots,n$, which has a purely
exponential time evolution. Thus $W^{(n)}$ of eq.~(\ref{g69b}b) is
distinguished from all other operators in $\CM_{z_R}^{(n)}$.

The microphysical decaying state operator associated with the $r$-th
order pole of the unitary S-matrix is according to its
definition~(\ref{g69a}b) a sum of the $W^{(n)}$. Because of the simple
form~(\ref{b82'}) (independence of the time evolution of $n$) this sum
has again a simple and exponential time evolution
\begin{equation}\label{b83}
	W(t)\equiv e^{-iH^\times t}We^{iHt}=e^{-\Gamma t}W\;;\quad t\geq0\,.
\end{equation}
Thus we have seen that the state operator which we conjecture from the
$r$-th order pole term describes a non-reducible ``mixed''
microphysical decaying state which obeys an exact exponential decay law.

We can return to the question that we started with when we set out
to conjecture the state operator for the (hypothetical) microphysical
state associated with the $r$-th order S-matrix pole: What is the
probability to register at the time $t$ the decay products
$|\psi^-\rangle\langle\psi^-|$ (or in general $\Lambda\equiv\sum_i
|\psi^-_i\rangle\langle\psi^-_i|$) if at $t=0$ the microphysical state
was given by $W$ of~(\ref{g69a}b)? From~(\ref{b83}) we obtain
\begin{equation}\label{b84}
	P_\Lambda(t)={\rm Tr}(\Lambda W(t))=e^{-\Gamma t}{\rm Tr}(\Lambda W)
	=e^{-\Gamma t}P_\Lambda(0)
\end{equation}
or in the special case of $\Lambda=|\psi^-\rangle\langle\psi^-|$:
\begin{equation}\label{b85}
	P_{\psi^-}(t)=\langle\psi^-|W(t)|\psi^-\rangle
		     =e^{-\Gamma t}\langle\psi^-|W|\psi^-\rangle
\end{equation}
This is exactly the same result as the result~(\ref{g62}b) for the
microphysical state $W^G$ associated with the first order pole of the
S-matrix and the result which is in agreement with the experiments on
the decay of quasistationary states. It is, however, important to note
that in our derivation of~(\ref{b82'}) and~(\ref{b85}) we proceeded in
a very specific order. We first derived~(\ref{b82'}) from~(\ref{b67}a)
and~(\ref{b67}b) and then calculated the matrix elements with
$\psi^-$ and not vice versa in order to avoid problems with the analyticity.


\section{Conclusion}\label{sec:conclusion}
\setcounter{equation}{0}

Vectors that possess all the properties that one needs
in order to describe a pure state of a resonance have been known for
two decades. These Gamow vectors $\psi^G$ are
eigenvectors of a self-adjoint Hamiltonian with complex eigenvalues
$E_R-i\Gamma/2$ (energy and width), they are associated with resonance
poles of the S-matrix, they evolve exponentially in time, and they
have a Breit-Wigner energy distribution. They also obey an exact
Golden Rule, which becomes the standard Golden Rule in the limit of
the Born approximation. The existence of these vectors in the rigged
Hilbert space allows us to interpret exponentially decaying 
resonances as autonomous microphysical systems, which one cannot do in
standard Hilbert space quantum mechanics. 

The mathematical procedure by which these Gamow vectors had been
introduced suggests a straightforward generalization to
higher order Gamow vectors which are derived from higher order
S-matrix poles. We have shown in this paper that the $r$-th order
pole of a unitary S-matrix leads to $r$ generalized eigenvectors
of order $k=0,~1,\cdots,~r-1$. These $k$-th order Gamow vectors are 
Jordan vectors of degree $(k+1)$ with complex eigenvalue
$E_R-i\Gamma/2$. They are basis elements of a generalized 
eigenvector expansion. But their time evolution has in 
addition to the exponential time dependence also a polynomial time
dependence, which is excluded experimentally. However, the generalized
eigenvector expansion suggests the definition of a state operator for 
microphysical decaying states of higher order. These state operators 
cannot be expressed as dyadic products of generalized vectors. 
But these state operators have a purely exponential time evolution.

There has been a lot of interest in the Jordan blocks for various
applications (see e.g.~\cite{lukierski,brandas,antoniou-tasaki,mondragon,stodolsky,antoniou-gadella,prigogine}).
Here it has been shown that Jordan blocks arise naturally from higher order
S-matrix poles and represent a self-adjoint Hamiltonian~\cite{22a} 
by a complex matrix in a finite
dimensional subspace contained in the rigged Hilbert space. Although
higher order S-matrix poles are not excluded theoretically, there has
been so far very little experimental evidence for their existence,
because they were always believed to have polynomial time dependence.
Since we have shown here that their non-reducible state operator
evolves purely exponential in time there is reason to hope that
these mathematically beautiful objects will have some application in
physics.

\vspace{24pt}\noindent
{\Large {\bf Acknowledgments}}

\vspace{12pt}\noindent
The inspiration for this paper came from three sources: Prof.~A. Mondragon's
talk at the 1994 workshop on Nonlinear, Deformed and Irreversible
Quantum Systems~\cite{mondragon}, which showed us the importance of
Jordan blocks; Profs.~I. Antoniou and M.~Gadella's unpublished
preprint of Spring 1995~\cite{antoniou-gadella}, 
which demonstrated that the Jordan
vectors are Gamow vectors of higher order S-matrix poles; and Prof.~I.
Prigogine's unrelenting talk of ``irreducible states''. We would like
to express our gratitude to them for valuable discussions and explanations.

\noindent This collaboration has been made possible by the financial
support of NATO.

\bibliographystyle{ieeetr}

\begin{thebibliography}{99}
\footnotesize{

\bibitem{doublepoles}
R.~G. Newton, 
  {\em Scattering Theory of Waves and Particles}, 2$^{\rm nd}$ ed.
  (Springer-Verlag, 1982); ~
M.~L. Goldberger and K.~M. Watson, 
  {\em Collision~Theory} (Wiley, New~York, 1964).

\bibitem{fonda-ghirardi-rimini}
Resonance states or unstable particles are associated with a complex
energy $E_R-i\Gamma/2$ and an exponential decay law, and in the
Hilbert space (due to its mathematical properties) there exists no
vector whose survival or decay amplitudes obey the exponential law
[L.~A. Khalfin, Sov.~Phys. JETP~{\bf 6}, 1053 (1958); ~ 
L.~Fonda, G.~C. Ghirardi, and A.~Rimini, 
  \newblock Rep. on Prog. in Phys.~{\bf 41}, 587 (1978),
  \newblock and references thereof].
The prediction of a small deviation from the exponential law by itself
would not constitute a severe discrepancy, since the exponential law
is only verifyable experimentally up to statistical
fluctuations. However, in the Hilbert space formulation the time
evolution is given by a unitary group and is therefore reversible,
whereas the experimental decay probabilities $P(t)$ can only be
measured for $t\geq t_0$ where $t_0$ is the time at which the unstable
particle has been produced. Furthermore, the decay probability $P(t)$
of a Hilbert space vector $\phi(t)=e^{-iHt}\phi_0$ can be shown to be
zero for all $t$ unless it is non-zero for almost all $t$ (if $H$ is
positive and self-adjoint)
[G.~C. Hegerfeldt, Phys.~Rev.~Lett. {\bf 72}, 596 (1994)],
whereas experimentally the number of decay products is zero before the
time $t_0$.


\bibitem{polynomial}
R.~G. Newton~\cite{doublepoles} sect.~6.4; 
M.~L. Goldberger and K.~M. Watson~\cite{doublepoles} chap.~8;
M.~L. Goldberger and K.~M. Watson, Phys.~Rev. {\bf 136} B1472 (1964);
A.~Bohm~\cite{bohm}, chapter XVIII.6; ~ 
A.~S. Goldhaber, {\em Meson~Spectroscopy}, p.~297, 
  edited by C.~Baltay and A.~H. Rosenfeld
  \newblock (Benjamin, New York, 1968).
This polynomial time dependence associated with higher order S-matrix
poles is of the order of $1/\Gamma$ and is {\em not} to be confused
with the non-exponential time dependence mentioned
in~\cite{fonda-ghirardi-rimini}  which was an artifact of the Hilbert
space mathematics and could therefore be made arbitrarily small by a
suitable choice of the Hilbert space vector.

\bibitem{bohm}
A.~Bohm, {\em Quantum~Mechanics}, {$1^{st}$}~ed.
  \newblock (Springer-Verlag,~Berlin, 1979),
  \newblock {$3^{rd}$}~ed. (1993).

\bibitem{bohm-group}
A.~Bohm in {\em Group Theoretical Methods in Physics}, Lecture Notes in
  Physics {\bf 94}, 245 (Springer-Verlag, Berlin, 1978); ~
A.~Bohm, Lett.~Math.~Phys.~{\bf 3}, 455 (1979);
A.~Bohm, J.~Math.~Phys.~{\bf 21}, 1040 (1981); ~
A.~Bohm, J.~Math.~Phys.~{\bf 22}, 2813 (1981).

\bibitem{gadella1983}
M.~Gadella, J.~Math.~Phys. {\bf 24}, 1462 (1983); ~
M.~Gadella, J.~Math.~Phys.~{\bf 24}, 2142 (1983); ~
M.~Gadella, J.~Math.~Phys.~{\bf 25}, 2481 (1984).

\bibitem{bohm-gadella}
A.~Bohm and M.~Gadella, {\em Dirac Kets, Gamow Vectors and Gel'fand Triplets}.
  \newblock (Springer-Verlag, Berlin, 1989).
 
\bibitem{22a}
Our Hamiltonians are always bounded from below, $H\geq0$, and
self-adjoint, which precisely means that the (adjoint of
$H$) $\equiv H^\dagger=\bar{H}\equiv$ (closure of $H$) in ${\cal H}$. By $H$ we
usually denote the operator in $\Phi$ which should then 
precisely be called essentially self-adjoint. The conjugate operator
(i.e. the extension of the adjoint $H^\times\supset H^\dagger\supset
H$ in $\Phi^\times\supset{\cal H}\supset\Phi$ has eigenkets in
$\Phi^\times$ which can have complex eigenvalues.

\bibitem{lee}
T.~D.~Lee, 
  {\em Particle Physics and Introduction to Field Theory}, chap. 15
  \newblock (Harwood Acad., Chur, 1981).

\bibitem{baumgartel-gantmacher-lancaster}
For definitions of Jordan operators and their properties, see
H.~Baumg{\"{a}}rtel, {\em Analytic Perturbation Theory for Matrices and
  Operators}, Chap.~2 (Akad. Verl., Berlin, 1984); ~
T.~Kato, {\em Perturbation Theory for Linear Operators} 
  (Springer-Verlag, Berlin, 1966); ~
F. R. Gantmacher, {\em Theory of Matrices}, sect.~VII.7
  (Chelsea, New York, 1959). 
P. Lancaster and M. Tismenetsky, 
  {\em Theory of Matrices}, $2^{nd}$ ed.,
  (Acad. Press, 1985).

\bibitem{lukierski}
J.~Lukierski, Bul.~Acad. Polish Science~{\bf 15}, 223 (1967); ~
Y.~Dothan and D.~Horn, Phys. Rev.~D~{\bf 1}, 6 (1970); ~ 
E.~Katznelson, J.~Math.~Phys.~{\bf 21}, 1393 (1980); ~
G.~Bhamathi and E.~C.~G. Sudarshan, Texas preprint (1995).

\bibitem{brandas}
E.~J. Br{\"{a}}ndas and C.~A. Chatzidimitriou-Dreismann in {\em Resonances},
  Lecture Notes in Physics {\bf 325}, 480, 
  edited by E.~J. Br{\"{a}}ndas and N.~Elander
  (Springer-Verlag, Berlin, 1987).

\bibitem{antoniou-tasaki}
I.~Antoniou and S.~Tasaki, Int.~J.~Quantum Chem.~{\bf 46}, 425 (1993).

\bibitem{mondragon}
A.~Mondrag\'{o}n, Phys.~Lett.~B {\bf 326}, 1 (1994); ~
E.~Hern\'{a}ndez and A.~Mondrag\'{o}n, 
  J.~Phys.~A Math. Gen. {\bf 26}, 5595 (1993); ~
A.~Mondrag\'{o}n {\em et~al.} in 
  {\em Nonlinear, Deformed, and Irreversible Quantum Systems}, p.~303,
  edited by H.~D. Doebner {\em et~al.}
  (World Scientific, Singapore, 1995); ~
E.~Hern\'{a}ndez, A.~J\'{a}uregui, and A.~Mondrag\'{o}n, 
  Rev.~Mex.~Fis. {\bf 38}, Supl. 2, 128 (1992).

\bibitem{stodolsky}
L.~Stodolsky in {\em Experimental Meson Spectroscopy}, p.~395, edited
by C.~Baltay and A.~H. Rosenfeld (Columbia Univ. Press, New York,
1970).

\bibitem{prigogine}
I.~Prigogine, Phys.~Rep. {\bf 219}, 93 (1992); ~
I.~Antoniou, Nature {\bf 338}, 210 (1989); ~
T.~Petrosky, I.~Prigogine, and S.~Tasaki, 
  Physica~A {\bf 173}, 175 (1991); ~
I.~Antoniou, I.~Prigogine, Physica A {\bf 192}, 443 (1993).


\bibitem{wilczek}
F.~Wilczek and A.~Shapere
  {\em Geometric Phases in Physics}
  \newblock (World Scientific, Singapore, 1989); ~
C.~A. Mead, Rev.~Mod.~Phys. {\bf 64}, 51 (1992); ~
A.~Bohm in {\em Integrable Systems, Quantum Groups, 
  and Quantum Field Theories}, p.~347, 
  edited by L.~A. Ibort and M.~A. Rodriguez 
  (Kluwer Academic, 1993); ~
ref~\cite{bohm}, chap.~XXII; ~ 
M.~V. Berry and J.~M. Robbins, 
  Proc. R. Soc. Lond.~A {\bf 442}, 659 (1993).

\bibitem{verbaarschot}
J.~J.~M. Verbaarschot, H.~A. Weidenm{\"{u}}ller, and M.~R. Zirnbauer,
  Phys.~Rep. {\bf 129}, 367 (1985).

\bibitem{antoniou-gadella}
I.~Antoniou and M.~Gadella, preprint (Intern. Solvay Institute,
Brussels, 1995). Results of this preprint were published in:
A.~Bohm et al., Rep.~Math.~Phys.~{\bf 36}, 245 (1995).

\bibitem{eigenvector}
As is the custom in the mathematics literature, the term generalized
eigenvector is used for two different things, which happen to be
related. In the rigged Hilbert space theory, a generalized eigenvector is
an element of $\Phi^\times$ which fulfills the
equation~(\ref{bohm23}). In the theory of finite dimensional
matrices or operators~\cite{baumgartel-gantmacher-lancaster}, 
a generalized eigenvector of a finite
dimensional operator $A$ belonging to eigenvalue $z$ is a vector
$|z\}$ which fulfills $(A-z)^m|z\}=0$
for a natural number $m$. From this definition we see that, according
to equation~(\ref{b58}'), the generalized vectors in the rigged
Hilbert space sense, $|z^-_R\mer^{(k)}\in\Phi^\times$, are also
generalized eigenvectors of the finite dimensional operator $H^\times$
restricted to the finite dimensional space ${\cal M}_{z_R}$
of~(\ref{bohm48}).

\bibitem{s-matrix}
see~\cite{doublepoles}, or~\cite{bohm} sect. XXI.4.

\bibitem{hogreve}
H.~Hogreve, Phys.~Lett.~A~{\bf 201}, 111 (1995).

\bibitem{cologne}
A.~Bohm in {\em Symp. on the Found. of Mod. Phys.}, Cologne 1993,
  p.~77, edited by P.~Busch, P.~Lahti, and P.~Mittelstaedt (World
  Scientific, Singapore, 1993); ~
A.~Bohm, I.~Antoniou, and P.~Kielanowski,
  Phys.~Lett.~A {\bf 189}, 442 (1994); ~
A.~Bohm, I.~Antoniou, and P.~Kielanowski,
  J.~Math.~Phys.~{\bf 36}, 2593 (1995).

\bibitem{fussnote}
The operators $\Om^+$ and $\Om^-$ are the M{\o}ller wave operators.  The 
Lippmann-Schwinger equation relates the (known) eigenvectors 
of the free Hamiltonian $K$ to two sets of eigenvectors of the 
exact Hamiltonian $H$:
\begin{equation}\nonumber
	|E^\pm \rangle=|E\,\rangle\;+\;{1\over{E-H\pm i\epsilon}}V\,
   	|E\,\rangle=\Om^\pm\,|E\,\rangle
\end{equation}
where $K\,|E\,\rangle =E\,|E\,\rangle$ and $H\,|E^\pm\rangle=E\,|E^\pm\rangle$
This defines the exact energy wavefunctions in terms of the 
in-- and out--energy wave functions, whose modulus gives the 
energy resolution of the experimental apparatuses.

\bibitem{duren-hoffman}
P.~L. Duren, {\em Theory of ${\cal H}^p$ Spaces}
  \newblock (Acad. Press, New~York, 1970); ~
Hoffman, K., {\em Banach Spaces of Analytic Functions} 
  (Prentice-Hall, Englewood Cliffs, N.~J., 1962). 


\bibitem{vanwinter}
C.~van Winter, Trans. Am. Math. Soc.~{\bf 162}, 103 (1971); ~ 
C.~van Winter, J.~Math. Anal. and Appl.~{\bf 47}, 633 (1974).

\bibitem{lee-oehme-yang}
T.~D. Lee, R.~Oehme, and C.~N. Yang, Phys. Rev.~{\bf 106}, 340 (1957).

\bibitem{nuss}
ref.~\cite{bohm}, sect. XVIII.5 and ref. thereof; ~~~
H.~M. Nussenzveig, 
  \newblock {\em Causality and Dispersion Relations} 
  \newblock (Acad.~Press, New York, 1972); ~~~
J.~R. Taylor
  \newblock {\em Scattering Theory}
  \newblock (Wiley, New York, 1972).

\bibitem{bohm-gadella-maxson}
A.~Bohm, S.~Maxson, M. Loewe, and M.~Gadella (to appear in Physica A).

\bibitem{ludwig-foundations}
G.~Ludwig, {\em Foundations of Quantum Mechanics}, 
  \newblock Vol.~I (Springer-Verlag, Berlin, 1983)
  \newblock Vol.~II (1985); ~
G.~Ludwig, {\em An Axiomatic Basis for Quantum Mechanics},
  \newblock Vol.~I (Springer-Verlag, Berlin, 1985), Vol.~II (1987).

}
\end{thebibliography}

\end{document}